\documentclass[namedreferences]{SolarPhysics}
\usepackage[optionalrh,solaenum]{spr-sola-addons} 
\usepackage{graphicx}        
\usepackage{color}           
\usepackage{url}             

\begin{document}

\begin{article}

\begin{opening}

\title{Multi-Wavelength Observations of a Flux Rope Failed in the Eruption and Associated M-Class Flare from NOAA AR 11045}
\author{Pankaj~\surname{Kumar}$^{1,2}$\sep
        A.K.~\surname{Srivastava}$^{1,4}$\sep
	B.~\surname{Filippov}$^{3}$\sep
        R.~\surname{Erd\'elyi}$^{4}$\sep
        Wahab~\surname{Uddin}$^{1}$
       }
\runningauthor{P.Kumar et al.}
\runningtitle{Failed flux rope and solar flare from AR NOAA 11045}
\institute{$^{1}$ Aryabhatta Research Institute of Observational Sciences (ARIES), Manora Peak, Nainital-263129, India.
		email: \url{aks@aries.res.in}}
\institute{$^{2}$Korea Astronomy and Space Science Institute (KASI), Daejeon, 305-348, Republic of Korea.
                     email: \url{pankaj@kasi.re.kr}}
\institute{$^{3}$ Pushkov Institute of Terrestrial Magnetism, Ionosphere and Radio Wave Propagation, Russian Academy of
       Sciences, Troitsk Moscow Region 142190, Russia.}
\institute{$^{4}$ Solar Physics and Space Plasma Research Centre (SP2RC), School of Mathematics and Statistics, The University of Sheffield, Sheffield, U.K.}
\begin{abstract}
We present the multi-wavelength observations of a flux rope that was trying to erupt 
from NOAA AR 11045 and the associated M-class solar flare on 12 February 2010 using space and ground based observations from TRACE, STEREO, SOHO/MDI, {\it Hinode}/XRT and BBSO.  While the flux rope was rising from the active region, an M1.1/2F class flare was triggered nearby one of its footpoints. We suggest that the flare triggering was due to the reconnection of a rising flux rope with the surrounding low-lying magnetic loops. The flux rope reached a projected height of $\approx0.15 R_{\odot}$ with a speed of $\approx$90 km s$^{-1}$ while the soft X-ray flux enhanced gradually during its rise. The flux rope was suppressed by an overlying field and the filled plasma moved towards the negative polarity field to the west of its activation site. We find the first observational evidence of the initial suppression of a flux rope due to a remnant filament visible both at chromospheric and coronal temperatures that evolved couple of days before at the same location in the active region. SOHO/MDI magnetograms show the emergence of a bipole $\approx$12 h prior to the flare initiation. The emerged negative polarity moved towards the flux rope activation site, and flare triggering near the photospheric polarity inversion line (PIL) took place. The motion of the negative polarity region towards PIL helped in the build-up of magnetic energy at the flare and flux rope activation site. This study provides a unique observational evidence of a rising flux rope that failed to erupt due to a remnant filament and overlying magnetic field, as well as associated triggering of an M-class flare.
\end{abstract}
\end{opening}

\section{Introduction}
Magnetic dynamo action in the solar interior continuously generates 
magnetic flux tubes that rise through the photosphere, add new 
magnetic flux systems to the photosphere, and finally fan out in 
the corona \cite{sch1999,asc2004}.  Such magnetic flux systems 
in the active regions are aligned with overlying arch filament systems, 
emerging in the form of $\Omega$/U-loops, or may exhibit 
``sea-serpent"--like shapes (\inlinecite{asc2004} and references therein). 
The newly emerging and growing magnetic fluxes may compel the topological 
changes in the overlying coronal magnetic fields, which 
may involve magnetic reconnection processes and their 
consequence in the form of eruptions at large and small
spatial-temporal scales. 
MHD simulations of such magnetic flux tubes have been performed 
in the subphotospheric layers \cite{fan2001,fan2003,fan2004} 
and also in the corona \cite{shibata1990}. 
Emergence of magnetic field and its reconnection with the
pre-existing field is not only responsible for the 
occurrence of large-scale eruptive phenomena, 
but it may also be the triggering mechanism for small-scale solar transients and explosive events 
(\opencite{roussev2001a},\citeyear{roussev2001b}, 
\citeyear{roussev2001c}). Recently, the emergence of 
magnetic flux rope from the convection zone to the corona 
and its interaction with the pre-existing fields have
been modelled and simulated by various authors ({\it e.g.}, \opencite{arch2004}, \citeyear{arch2005},
 \citeyear{arch2007}).
 
The emergence of a dipole is not a sufficient
condition to trigger a flare, because the emerging 
flux region may be too small or may have a non-favourable 
orientation \cite{martin1984,asc2004}. Pre-flare brightenings and 
flares probably may not have direct relevance with the emerging 
flux tubes in many cases. However, they may be associated with
the coupled energy release processes such as current 
driven plasma micro-instabilities, etc.\cite{karpen1986}. 
It is most likely that the magnetic flux emergence does not trigger 
flares directly, but may instead increase the magnetic complexity locally 
by adding new magnetic flux and, helicity, or by inducing  motions of sunspots. When 
this magnetic complexity crosses the critical threshold, flares and associated eruptions may be triggered \cite{asc2004,sri2010}. 
However, flare models that are directly driven by flux emergence have 
also been reported \cite{hey1977,asc2004}. Moreover, the rising process of  the magnetic field through the convection zone and its emergence through 
the photosphere, chromosphere and corona are not well understood. 
The exact reason of the activation of helicity and twist in the 
pre-existing magnetic flux ropes are also not well explored 
both in the observations and theory.

The magnetic flux from the sub-photospheric levels may also 
be associated with the helicity and twist, which leads to 
more stored magnetic energy and instabilities. Recently, 
more observational signatures became evident that one of the origins of this twist is at a near sub-photospheric level from which the flux tube swirls above the surface \cite{bonet2008,wede2009}. 
These phenomena are also described by recent theoretical models \cite{simon1997,shel2011a,shel2011b,fedun2011}.
The solar flares are mainly distinguished by two categories, {\it i.e.}, confined and 
eruptive flares, and both of them usually take place in rather complex morphology 
of overlying magnetic fields in the active regions. The helicity and
instability generated 
in such complex magnetic fields are the most likely cause of stored excess energy in the active regions that is later released in the form of solar 
flares and initiates the related plasma eruptions.  
The emergence of such twisted and unstable magnetic flux tubes, or their activation
and interactions in the 
active regions, may also trigger the flare energy release and
associated eruptions as studied recently by
,{\it e.g.}, \inlinecite{pankaj2010a}, \inlinecite{pankaj2010b}, \inlinecite{pankaj2010c}, \opencite{sri2010}. 
On the modelling side, most of the numerical simulations have used magnetic flux emergence 
and sunspot rotation caused by photospheric plasma flows to increase the 
magnetic helicity leading to the final eruption ({\it e.g.}, \opencite{amari2000},
\opencite{torok2003}, \citeyear{torok2005}).

\inlinecite{shen2011} have reported the dynamics of a filament that failed 
in the eruption several times, and finally succeeded to erupt with the 
triggering of a C-class flare process and associated CME on 8 February 
2010 from active region (AR) NOAA 11045.
Before that work, the
observational evidence of flux 
ropes failed in eruption has been reported as well \cite{ji2003,alex2006,liu2009}.
\inlinecite{boris2001} have studied the 
critical height and compared it with the stability criterion of the filaments. 
However, most of the previous observational studies of failed eruption 
could not reveal the exact mechanism associated with it. 
In this paper, we will study the dynamics of a flux rope in 
the AR 11045 on 12 February 2010 that moved up higher in the 
corona but still failed in the eruption. We also study the energy 
build-up and release processes of an M-class flare associated 
with the activation of this flux rope. We find the first 
observational signature of the rising flux rope that is 
failed due to an overlying remnant filament. 
In Section 2 we will present the multi-wavelength observations. 
We will discuss the magnetic configuration and scenario of the 
event in Section 3. Discussion and Conclusions will be presented 
in the last section.

\section{Observations}

\begin{figure}
\centerline{
\includegraphics[width=0.8\textwidth]{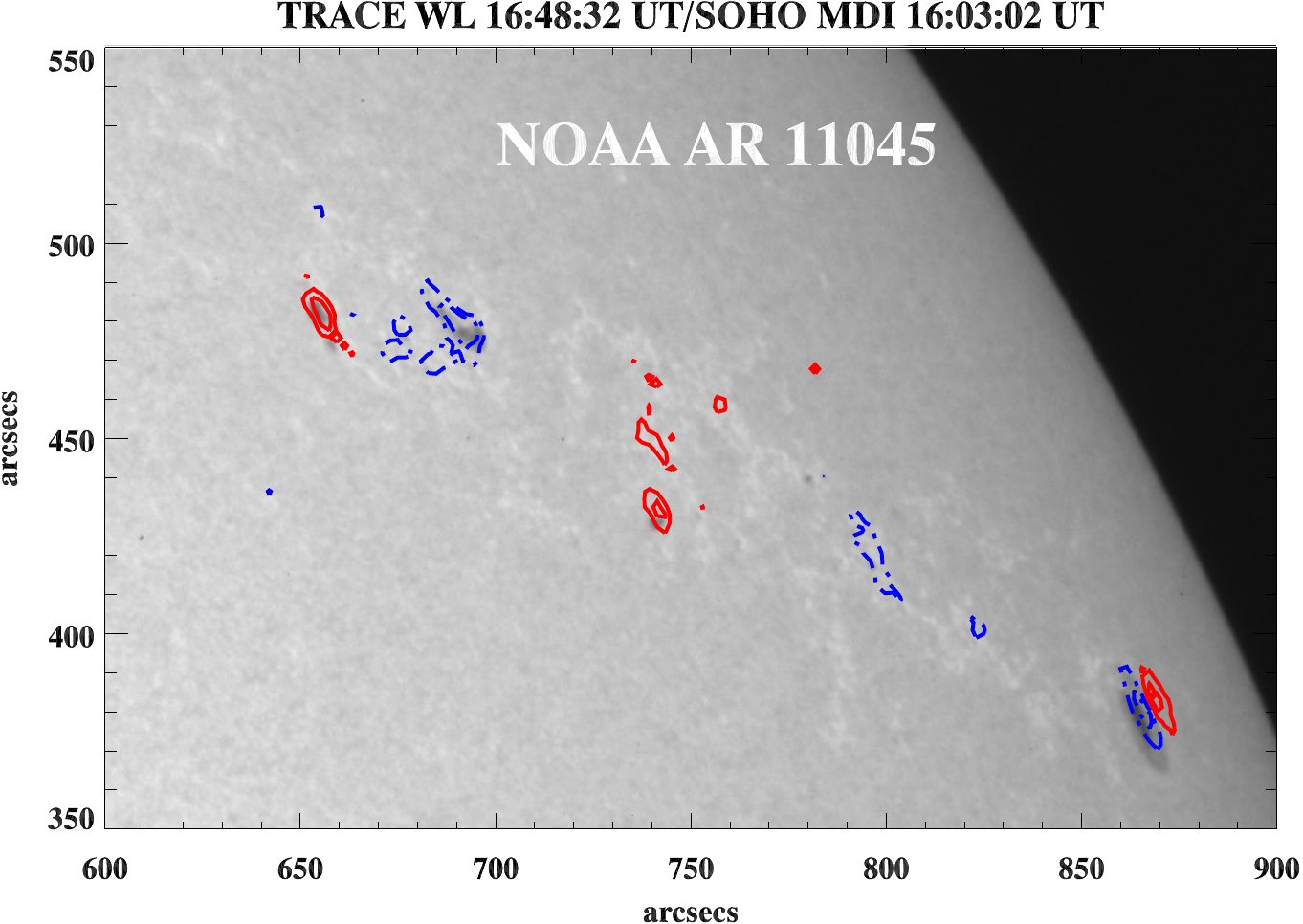}
\thicklines
$\color{yellow} \put(-197,125){\circle{200}}$
}
\caption{NOAA AR 11045 on 12 February 2010. Red (solid) 
contours show the positive polarity magnetic fields whereas blue 
(dashed) contours show the negative polarity field regions. The 
region shown by circle indicates the site of new flux emergence in the
vicinity of the flare and flux rope activation site. Solar north at the top, west to the right in this and all the following figures.}
\label{wl_mdi}
\end{figure}
The active region NOAA 11045 appeared to be very dynamic and produced more than 40 flares above C-class 
from 6 to 14 February 2010 \cite{xu2010} during its passage over the solar disk. NOAA 11045 continued to decay on 12 February 2010 and was located 
near the western limb ($\approx$N22,W66) showing a $\beta$ magnetic configuration (Figure \ref{wl_mdi}). However, this active region also showed a rapid and large flux emergence during its journey over the solar disk. According to  
GOES soft X-ray flux measurements the M-class flare on 12 February 2010 was started at $\approx$17:52 UT, reached a maximum at  $\approx$18:08 UT, and ended at  $\approx$18:15 UT (refer to Figure \ref{ht}). The flare was classified as 2F which initiated at  $\approx$17:59 UT and ended at  $\approx$18:41 UT in H$\alpha$. The flare plasma was gradually rising with the activation and upward motion of a rising flux rope at the same site, which finally failed in the eruption. In the following subsections we describe the multi-wavelength observations of a rising twisted flux rope and associated M-class flare.

\begin{figure}
\centerline{
\includegraphics[width=0.5\textwidth]{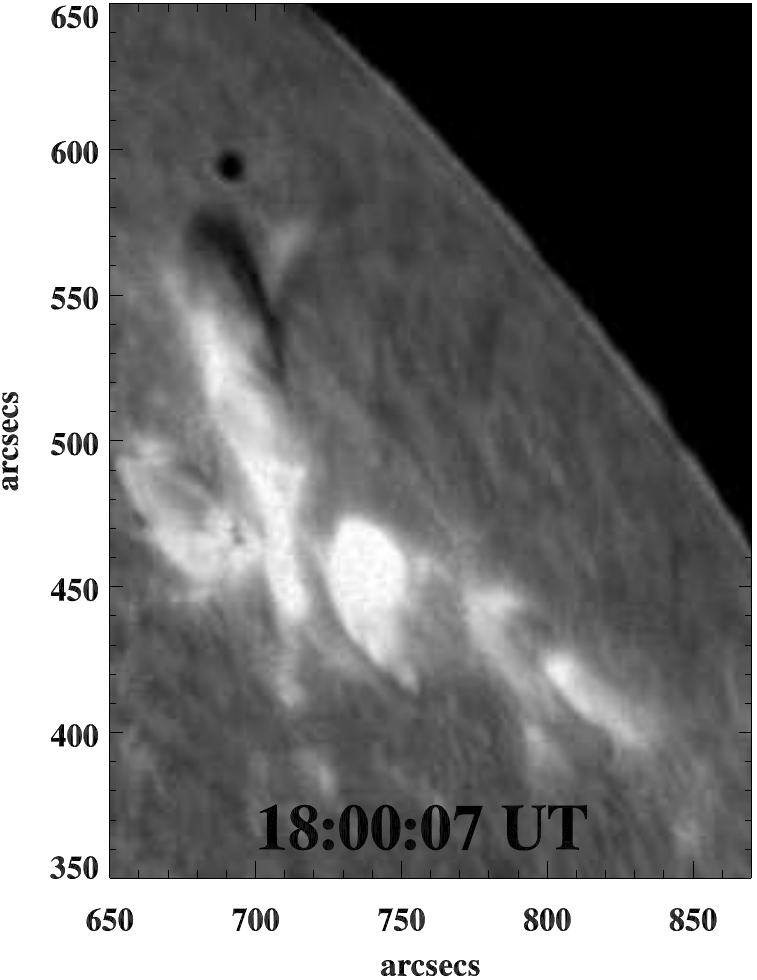}
\thicklines
$ \color{yellow} \put(-128,192){\vector(0,-1){15}} \put(-150,195){Remnant filament}$
$ \color{yellow} \put(-80,180){\vector(-1,0){15}} \put(-75,180){Rising flux}  \put(-75,170){rope}$
\includegraphics[width=0.5\textwidth]{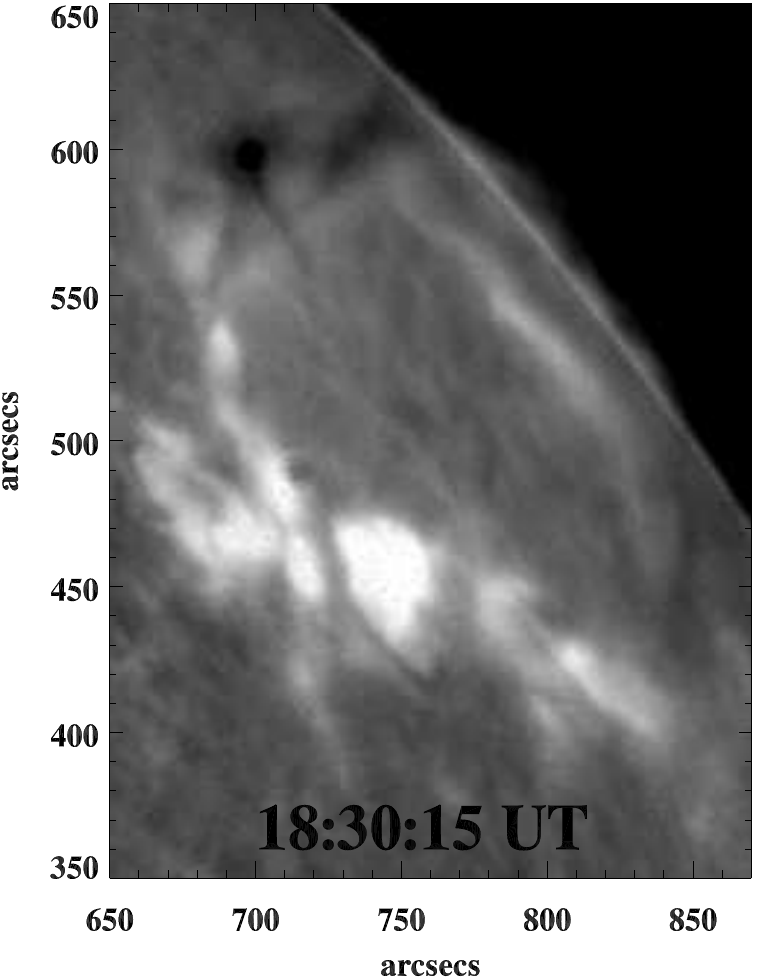}
$ \color{yellow} \put(-40,205){\vector(-1,-1){15}}$
}
\centerline{
\includegraphics[width=0.5\textwidth]{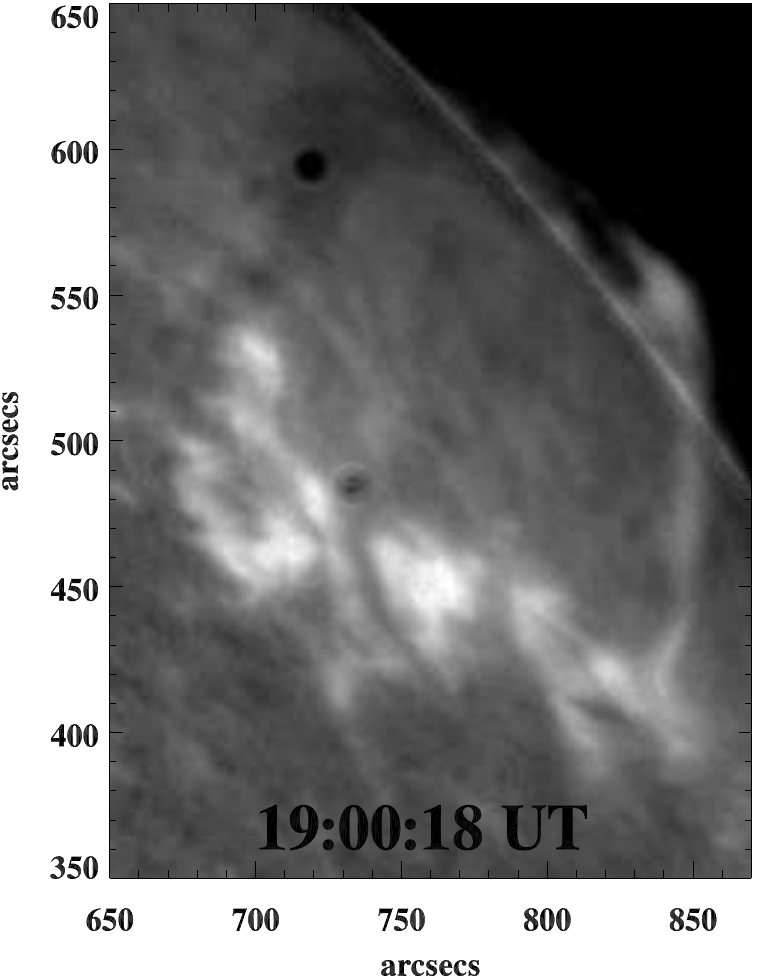}
\thicklines
$ \color{yellow} \put(-8,85){\vector(-1,-1){15}}$
\includegraphics[width=0.5\textwidth]{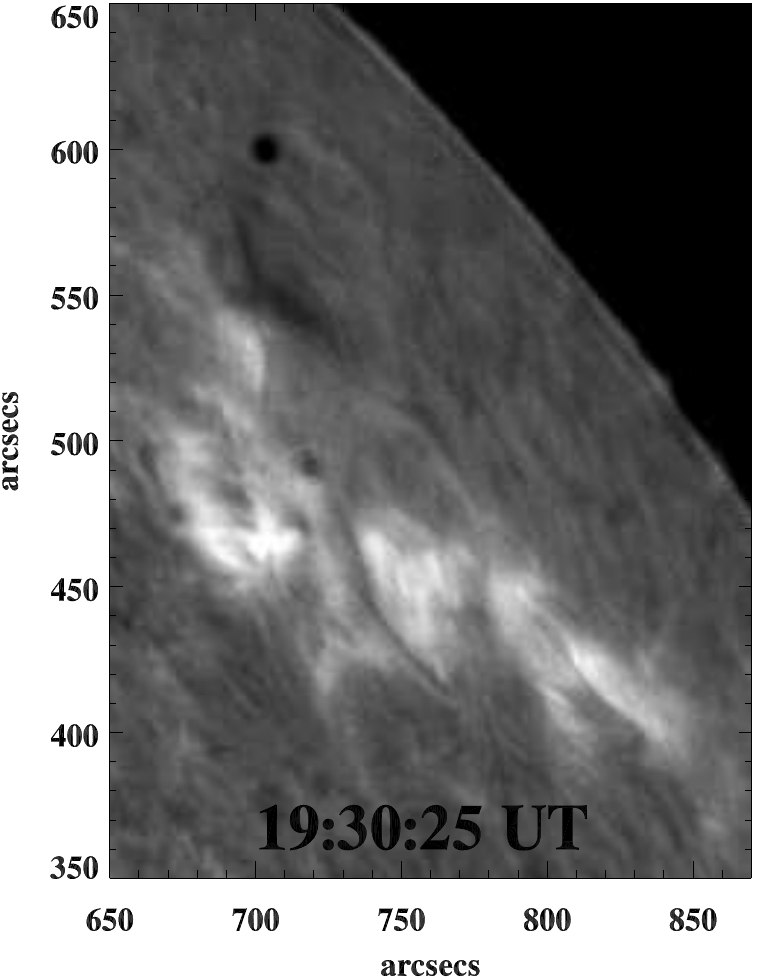}
}
\caption{H$\alpha$ image sequences from BBSO (Big Bear Solar Observatory) showing the 
activation of the flux rope, triggering an M1.1/2F flare (shown by yellow arrows) overlying the activated flux rope. In the top-left panel
the remnant filament in the form of a black cool arch is overlying the  
activated flux rope causing its suppression initially.}
\label{bbso}
\end{figure}

\begin{figure}
\centerline{
\includegraphics[width=0.5\textwidth]{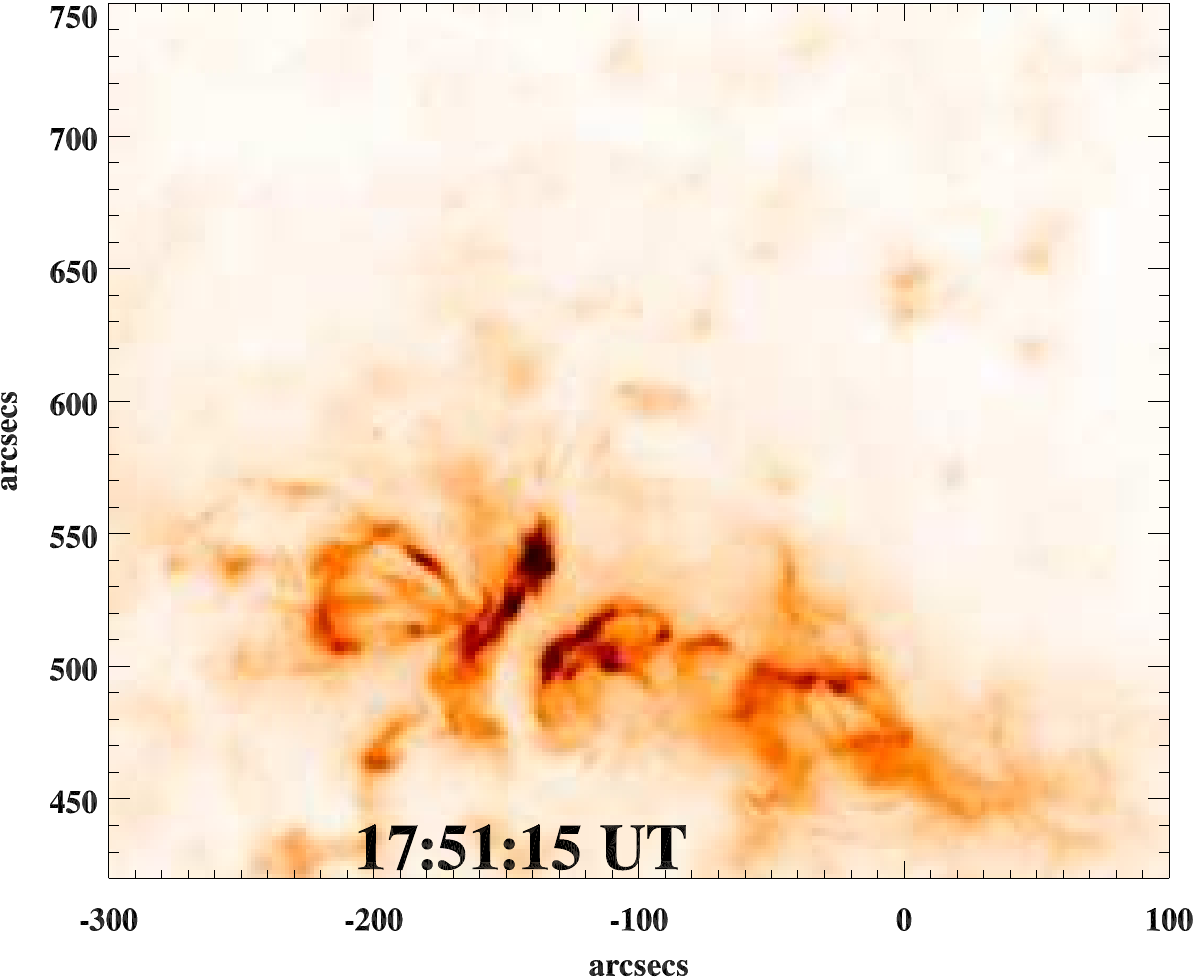}
\thicklines
$ \color{black} \put(-118,75){\vector(1,-1){15}}\put(-160,80){Twisted flux rope}$
\includegraphics[width=0.5\textwidth]{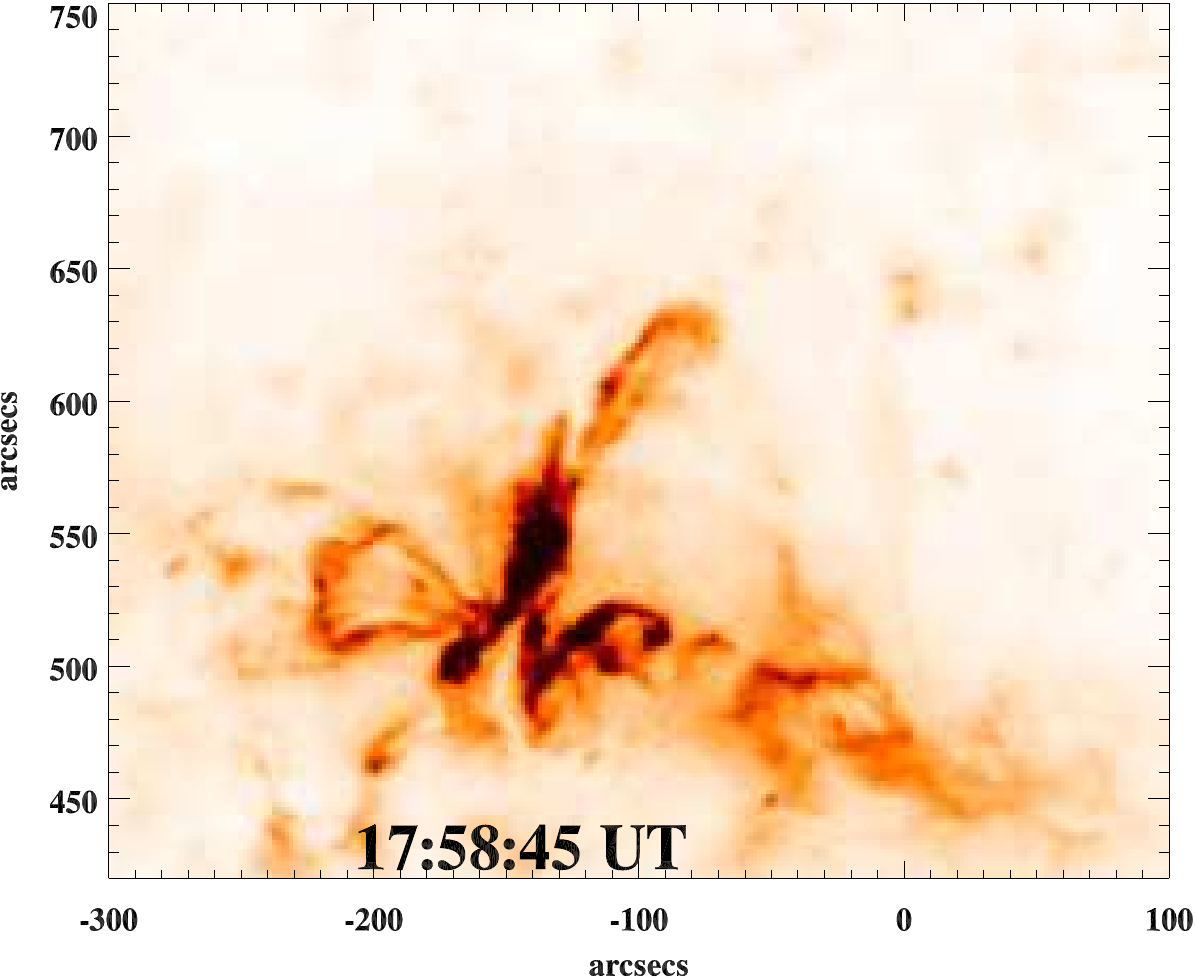}
$ \color{black} \put(-105,110){\vector(1,-1){15}}$
}
\centerline{
\includegraphics[width=0.5\textwidth]{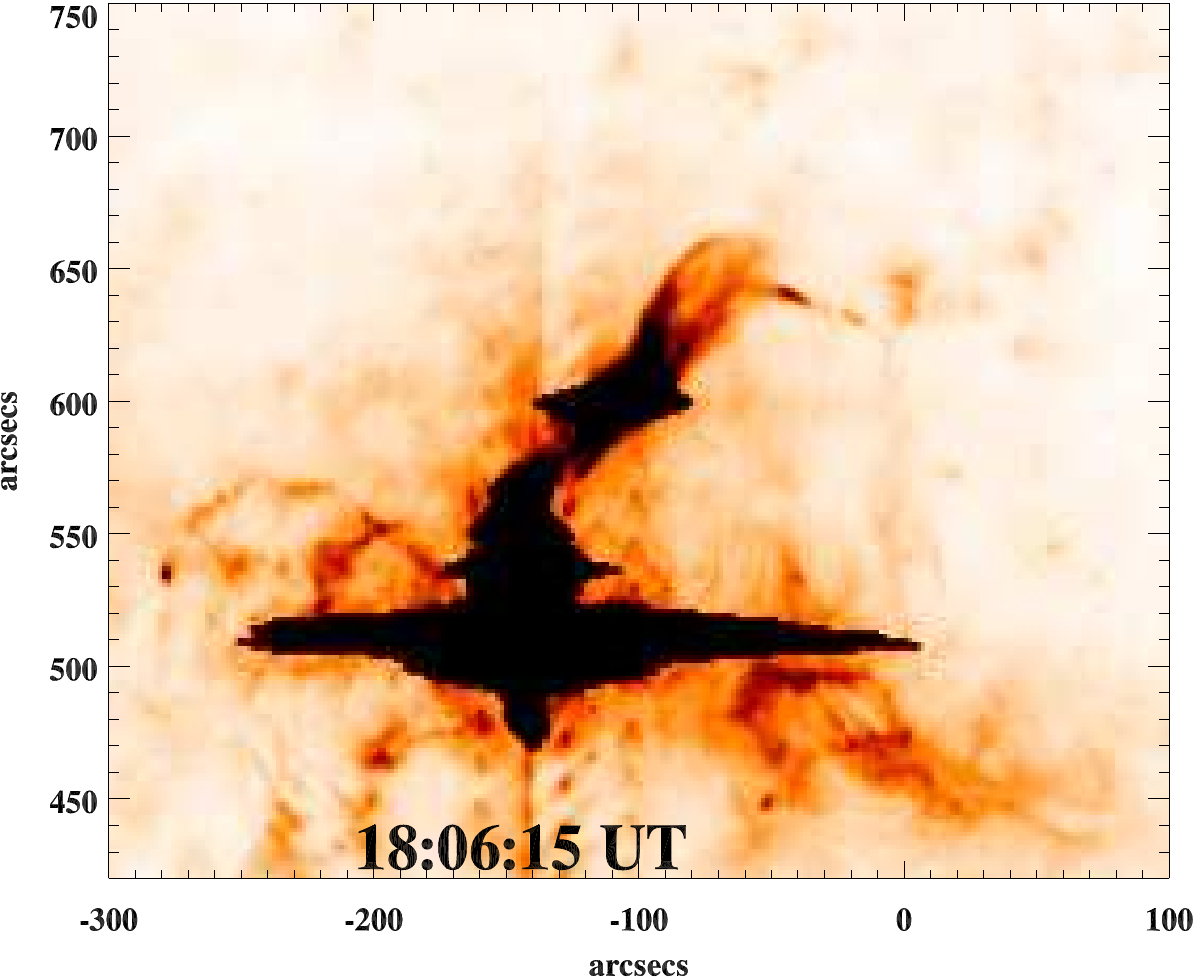}
\thicklines
$ \color{black} \put(-124,70){\vector(1,-1){15}}\put(-145,80){Flare}$
\includegraphics[width=0.5\textwidth]{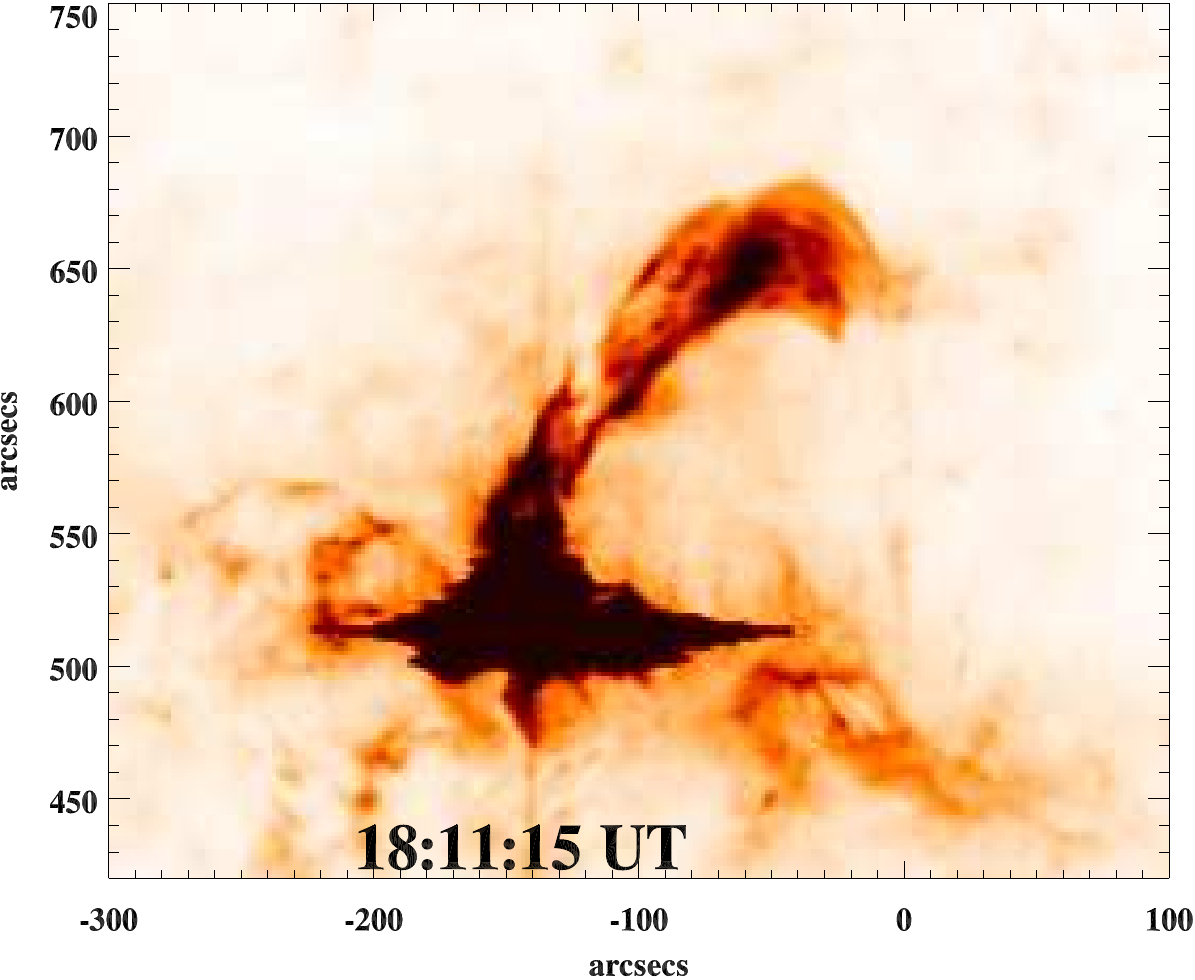}
}
\centerline{
\includegraphics[width=0.5\textwidth]{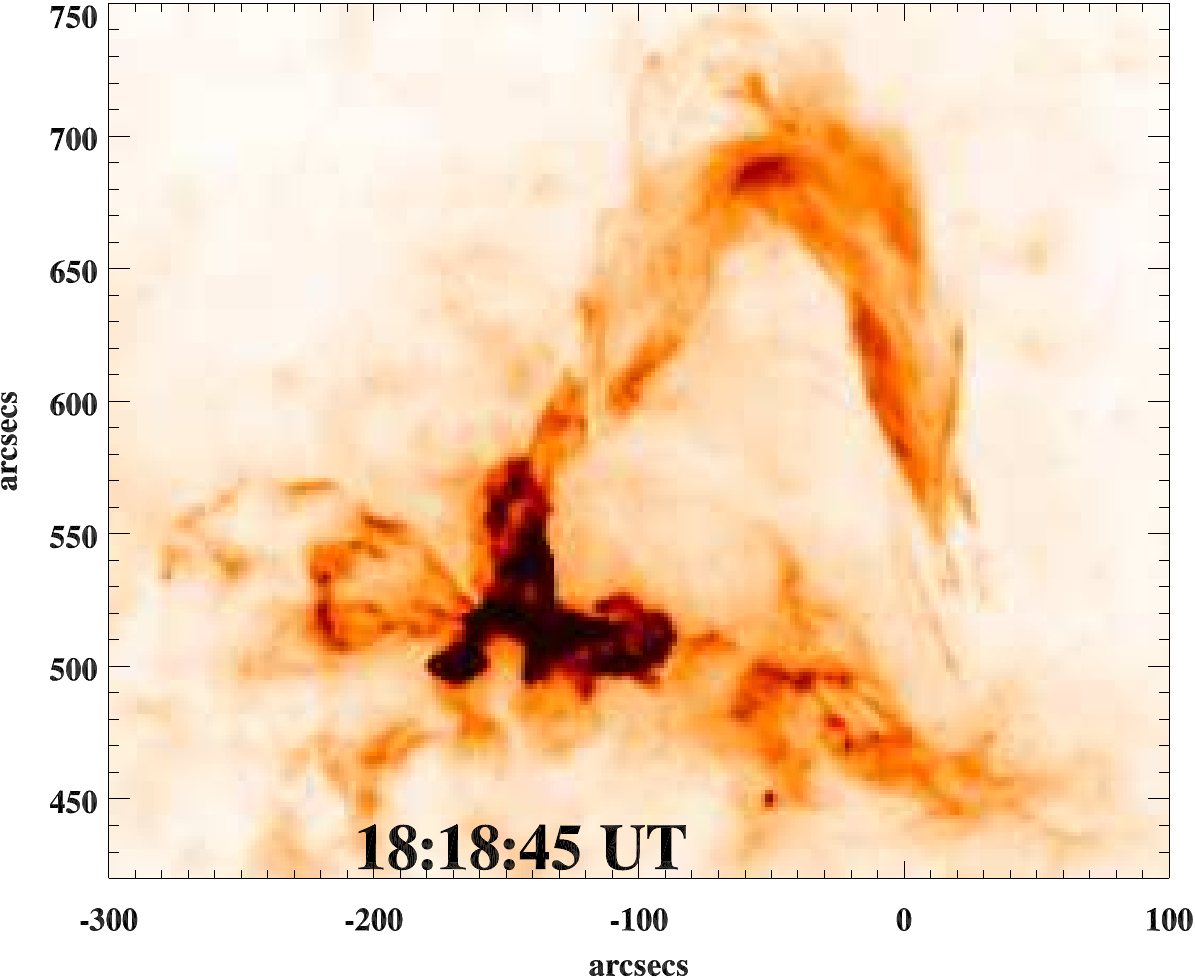}
\includegraphics[width=0.5\textwidth]{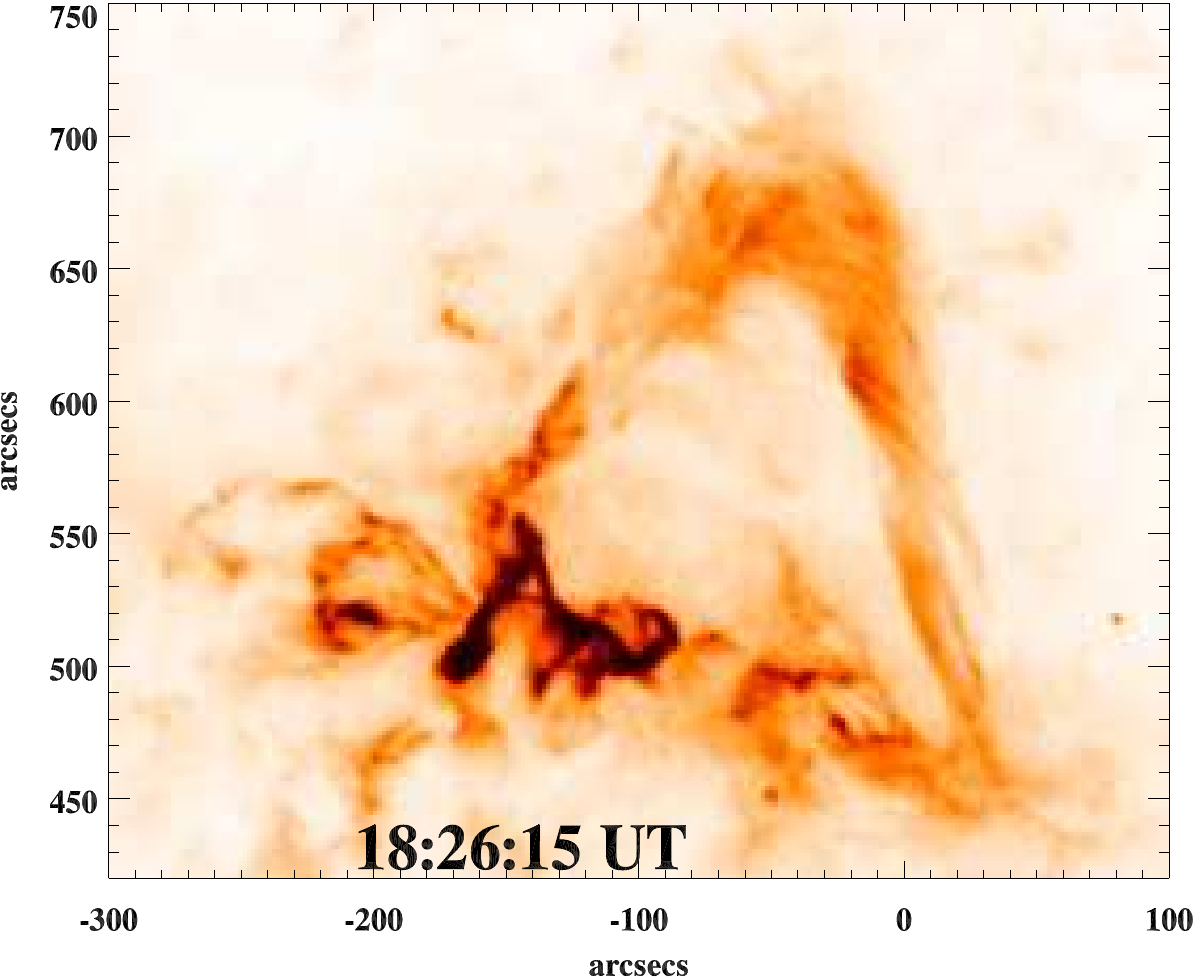}
}
\caption{STEREO A /SECCHI 304 \AA \ (He {\sc ii}) negative images showing the 
activation of a flux rope, and the triggering of the M1.1/2F flare 
in NOAA AR 11045 at the activation site nearby its eastern footpoint.}
\label{secchi_304}
\end{figure}

\begin{figure}
\centerline{
\includegraphics[width=0.8\textwidth]{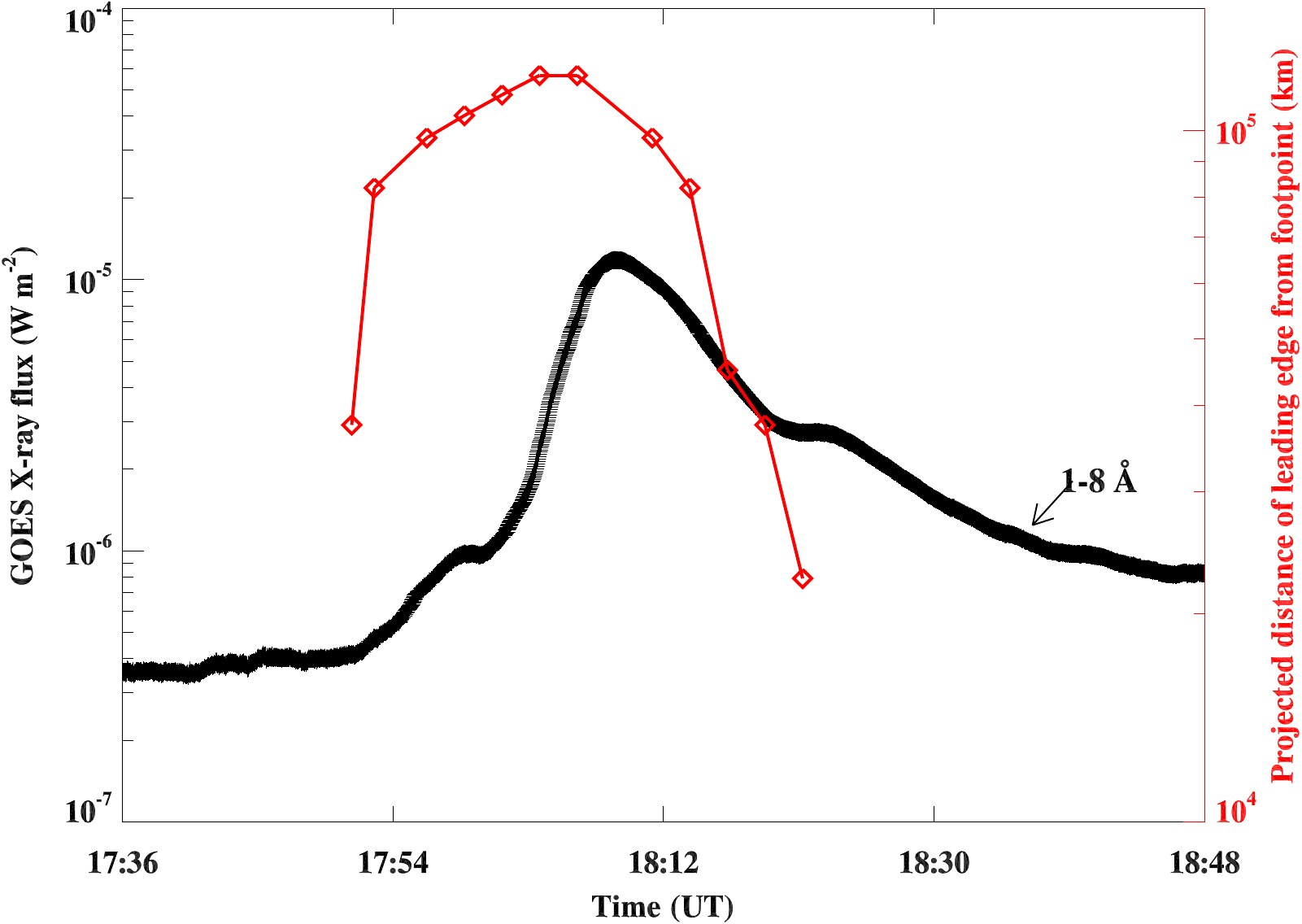}
}
\caption{GOES soft X-ray flux profile (thick black curve) and rise and fall motion of the hot
plasma (thin red curve). The latter is represented by location of the leading edge of the bright structure measured from the line connecting the flux rope footpoints.
}
\label{ht}
\end{figure}
\subsection{BBSO H$\alpha$ and STEREO Observations}
Figure \ref{bbso} displays selected H$\alpha$ images of the active region 
observed at Big Bear Solar Observatory (BBSO).
A careful investigation of the H$\alpha$ data reveals a rising bright twisted flux rope structure below the remnant dark filament (indicated by arrows). The H$\alpha$ images show a rising bright structure as part of the flux rope, and two ribbon flare nearby one of the footpoints at its activation site. 
However, we note that there was a small remnant filament 
structure lying above the rising flux rope. This part of the filament was visible before and after the activation and rise of the flux rope, and acted like a semi-circular cavity inside which the twisted flux rope was situated and confined initially. 
Here we note only that this moving structure is not a part of an existing filament. However, it can be a part of a complex flux rope containing both the remnant filament and the flux tubes that, were not filled with dense plasma initially. The bright
flux rope in H$\alpha$ indicates a high temperature of the structure compared to the cool quiescent filaments. Twists in the structure may also be noticed in these images, showing a transversal kink--like structure. 

Figure \ref{secchi_304} is a selected but typical sequence of EUV (304 \AA) STEREO-A
/SECCHI images. This wavelength corresponds to the emission from
 ionized helium (He {\sc ii}), which forms at temperatures above 80,000 K.  
The BBSO H$\alpha$ images represent the plasma formed at 10$^{4}$ K temperature in the chromosphere, while the STEREO He {\sc ii} 304 \AA\ images also reveal the upper chromosphere and lower transition region at greater temperature.
The NOAA AR 11045 was located near the center of the solar disc (in the eastern side) in the field of view of STEREO-A SECCHI.  
On the other hand, TRACE also observed NOAA AR 11045 near to the western limb. Therefore, the observations of the same active region from different angles provide the opportunity to view the structure from different view angles.
These observations show a rising plasma structure maintained at coronal temperature as indicated by arrows. Initially, the structure was visible at 17:51 UT associated with a flare initiation (refer to the first image). Then, the plasma associated with the flux rope started to move up. The flare was triggered nearby the eastern footpoint of the structure as it moved away from the active region. The flare maximum took place at $\approx$18:07 UT. The leading edge of the twisted structure reached the maximum distance from the footpoint, at the projected height of $\approx$150 Mm (see the panel at 18:11 UT on Figure 3). 
The flare brightening started to reduce as the structure started to bend towards the western side of the active region.  The leading edge of the flux rope moved up to the western part of the AR (see the panel at 18:26 UT on Figure 3). This snapshot shows the multiple twisted flux tubes in the flux rope structure. 
\begin{figure}
\centerline{
\includegraphics[width=0.5\textwidth]{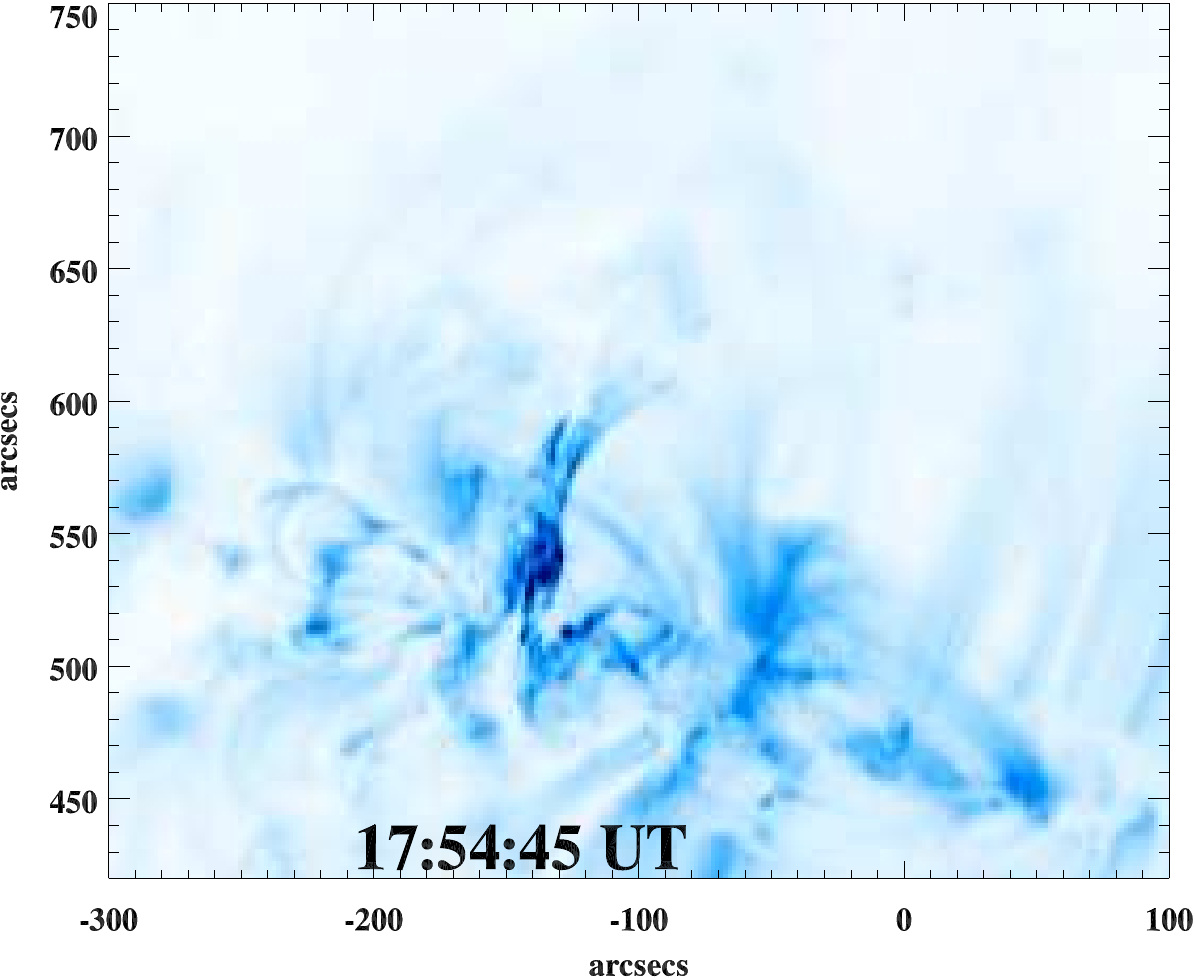}
\thicklines
$ \color{red} \put(-118,75){\vector(1,-1){15}}\put(-145,80){Flux rope}$
\includegraphics[width=0.5\textwidth]{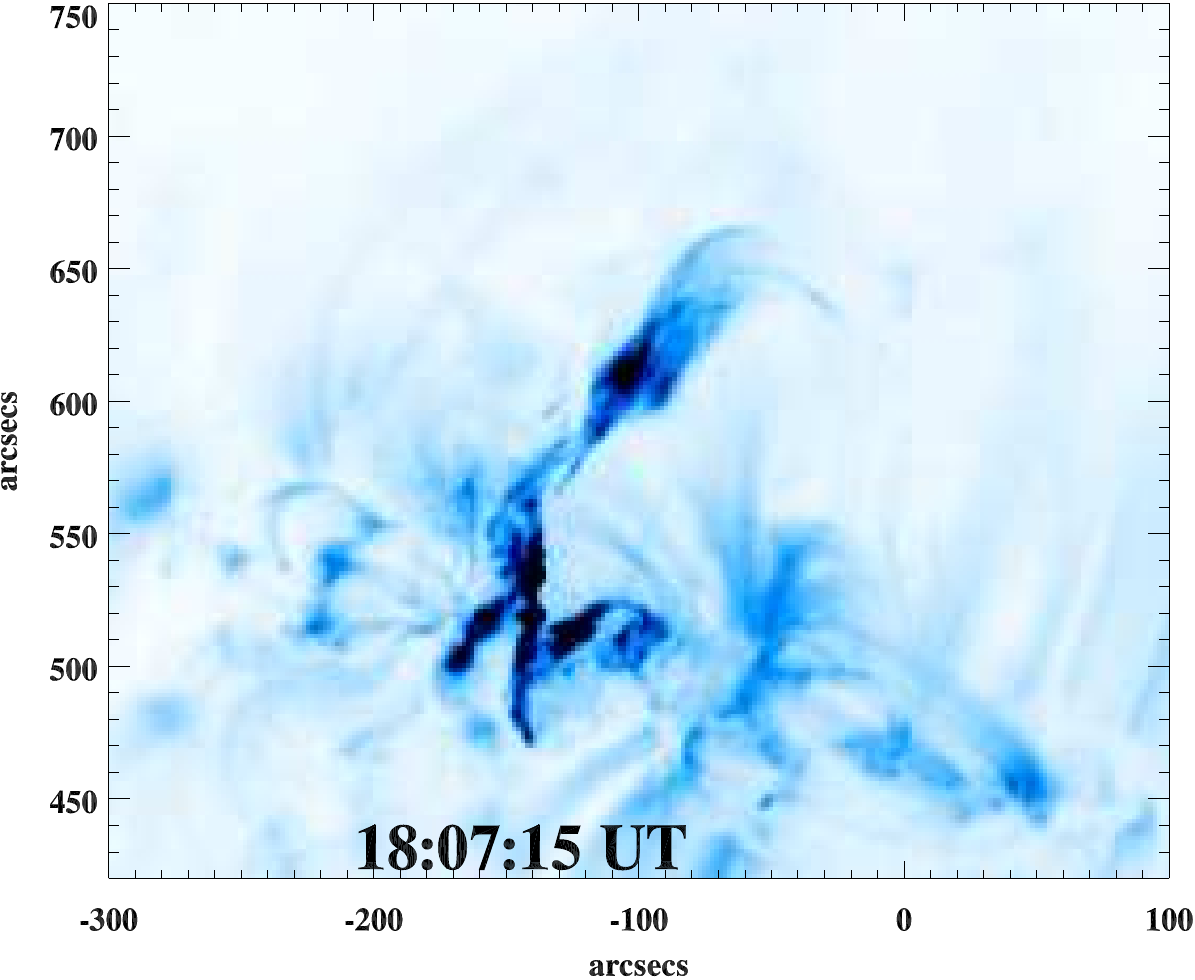}
}
\centerline{
\includegraphics[width=0.5\textwidth]{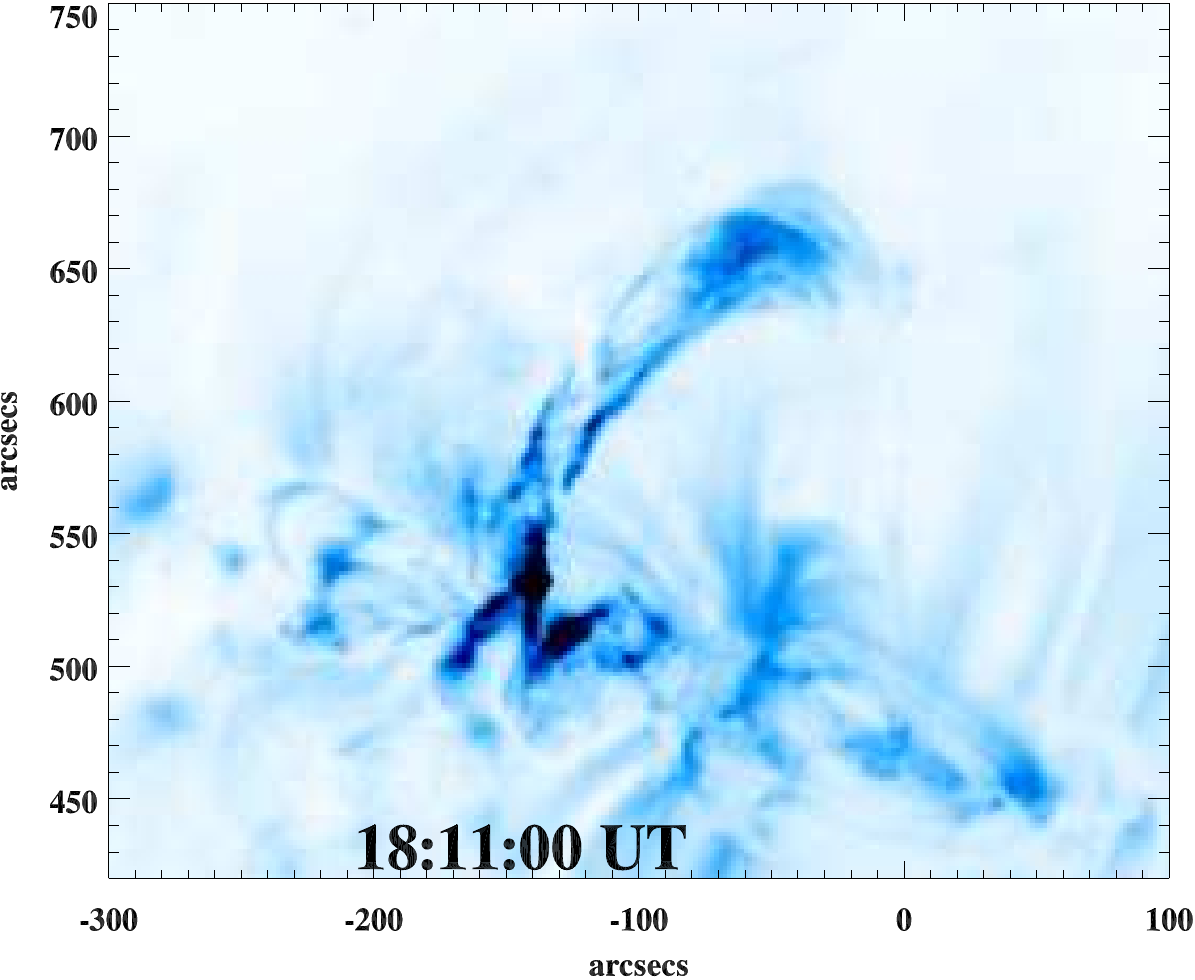}
\includegraphics[width=0.5\textwidth]{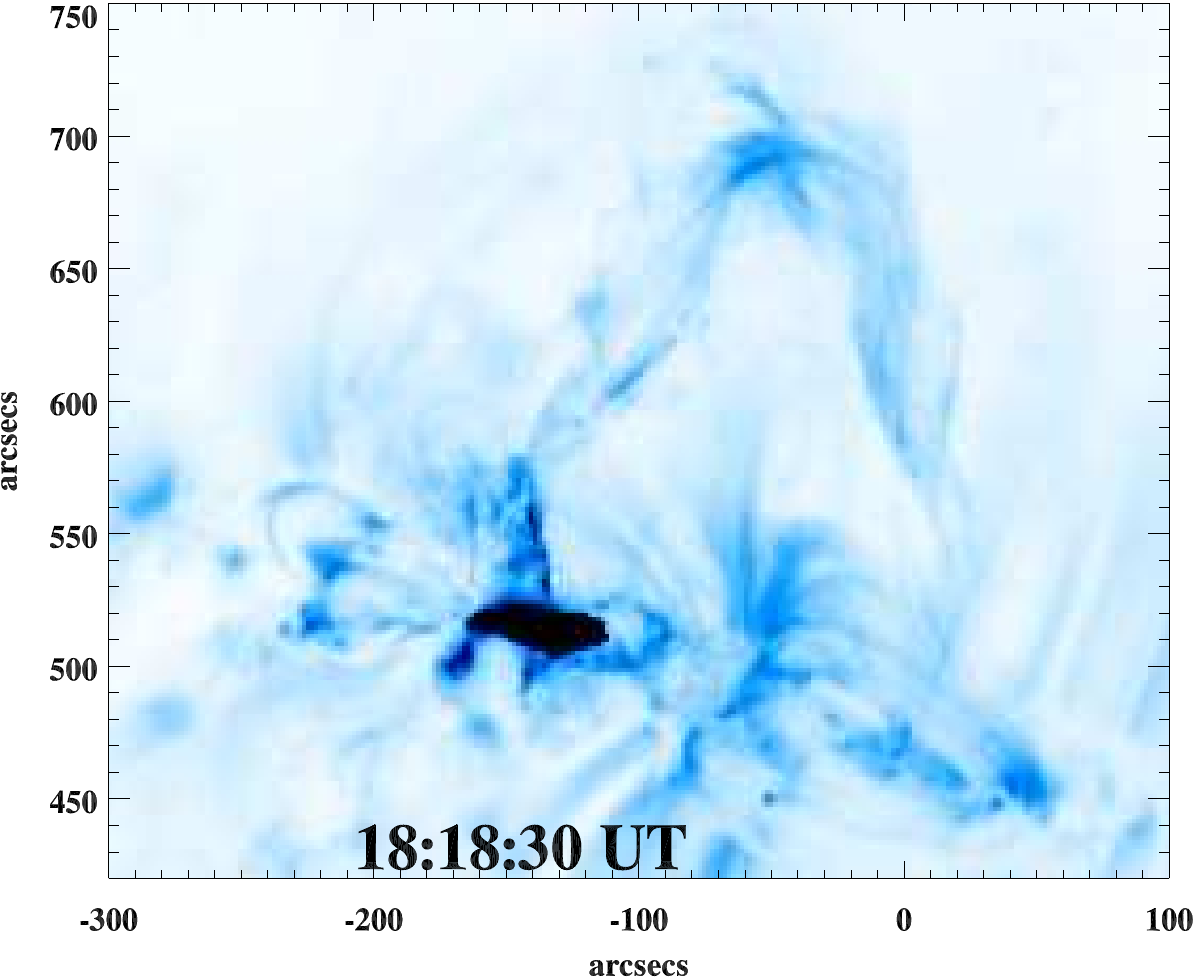}
}
\caption{STEREO A /SECCHI 171 \AA \ EUV negative image sequence showing the 
activation of a flux rope, and triggering of the M1.1/2F flare.}
\label{secchi_171}
\end{figure}

Figure \ref{ht} (red curve) shows the  projected distance vs. time profile of the
leading edge of the flux rope 
derived from the STEREO A/SECCHI 304 \AA \ image sequence. This is the lower bound to the estimated height, 
and the actual height may be much larger. This figure
reveals that the rising motion of the flux rope shows good correlation with 
the soft X-ray flux enhancement. This may be the most likely signature of a
magnetic reconnection between the flux rope and the surrounding lower 
loop-system. The reconnection process seems to progress with the rising motion of the flux rope and resulted in the slow rise in soft X-ray flux. The flux 
rope reached the maximum height of $\approx$1.2$\times$10$^{5}$ 
km and then its leading edge slowly moved down towards the active region 
 as a hook shaped structure, which was connected to the western side 
of the AR.  From the linear fit to the positions of the leading edge of the 
rising and bending flux rope, its ascending speed 
was estimated as $\approx$90 km s$^{-1}$, while the bending speed was $\approx$115  
km s$^{-1}$. These are the lower-bound projected speeds derived from the linear fit 
of the motion of the leading edge of the flux rope plasma in the EUV image sequence. 
Both these lower bound speeds show a slower 
evolution followed by suppression of the flux rope during the total span of the flare activity.

Figure \ref{secchi_171} depicts the NOAA AR 11045 observed by SECCHI EUV in the 171 \AA \ wavelength, 
which corresponds to the plasma temperature of 1.3 MK. In these data 
the same rising flux rope structure in coronal temperature can be clearly identified. Analysis of the data shows the magnetic topology of 
the active region connecting smaller and higher loop-systems. The small 
post-flare loop system near to the flare and flux rope activation site can be seen 
during the decay phase of the flare (refer to snapshots at 18:18 UT). These post-flare 
loop systems were formed after the flaring activity at the flux tube activation
site near the polarity inversion line (PIL). The chromospheric BBSO H$\alpha$ and STEREO/SECCHI He {\sc ii} (304 \AA) image sequences reveal 
the low-atmospheric scenario of the flux rope dynamics at its activation site near the PIL, its encounter with the remnant 
filament and subsequent suppression in the rising motion, and the brightening of flare ribbons as a signature of particle acceleration from the 
reconnection site. On the other hand, the STEREO/SECCHI 171 \AA\, TRACE and XRT observations provide the clues of the rising flux rope 
in the coronal field and its reconnection with nearly closed loop systems that released the stored flare energy.
 
\begin{figure}
\centerline{
\includegraphics[width=0.5\textwidth]{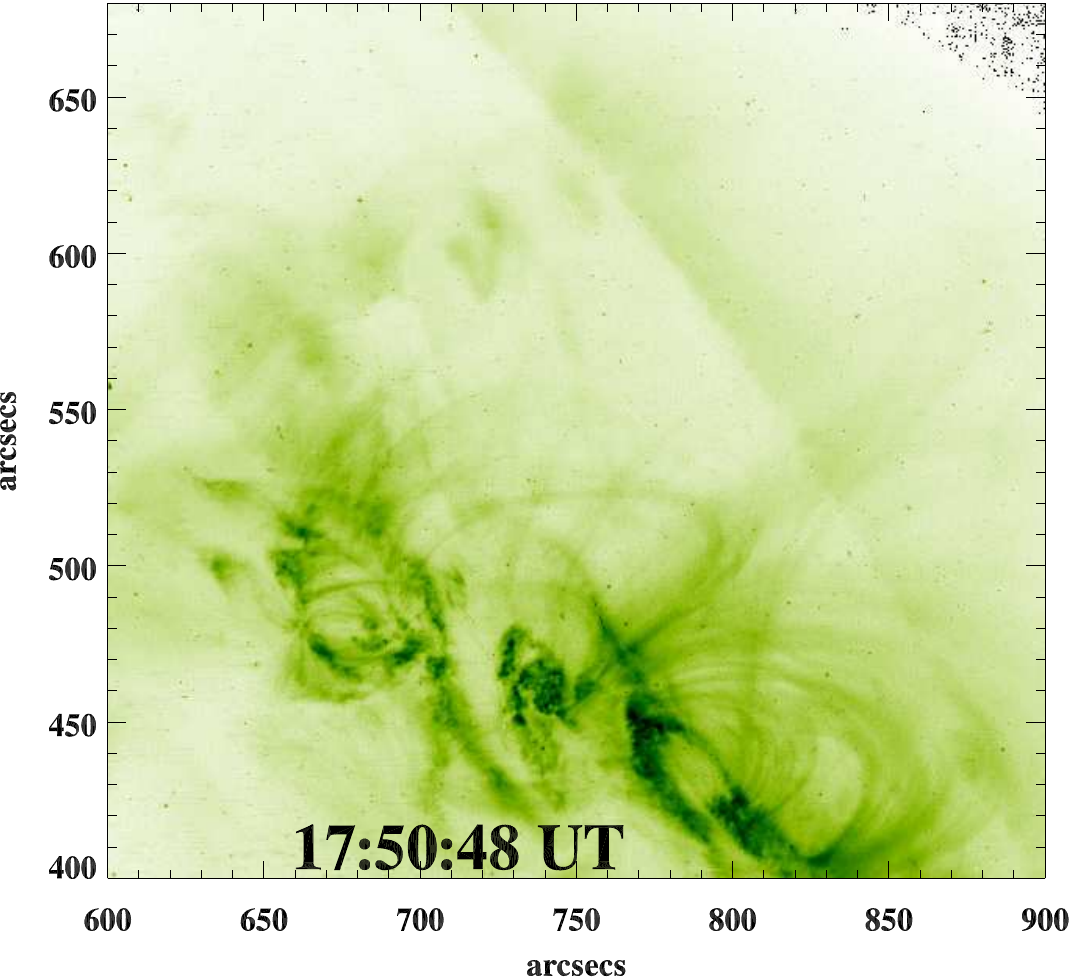}
\includegraphics[width=0.5\textwidth]{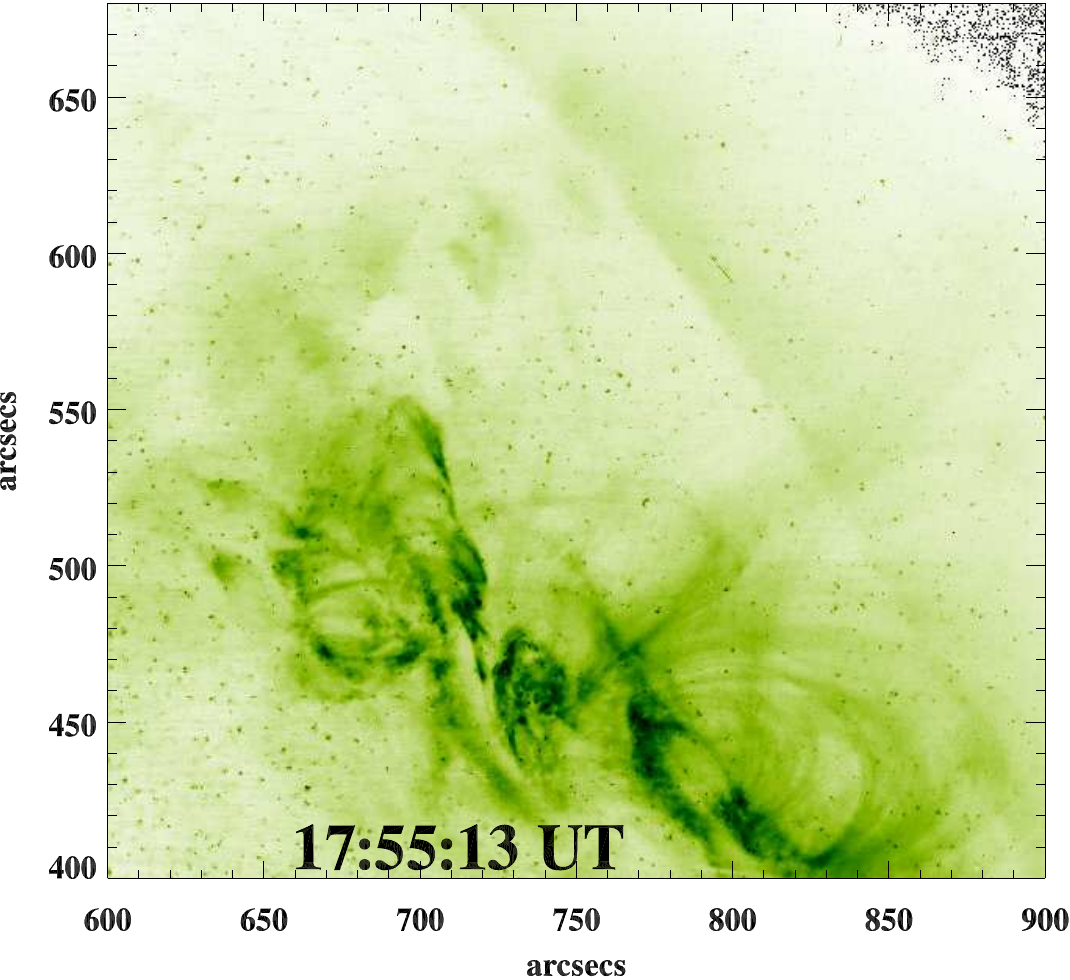}
\thicklines
$ \color{black} \put(-102,89){\vector(0,-2){15}}\put(-145,100){Twisted flux rope}$
}
\centerline{
\includegraphics[width=0.5\textwidth]{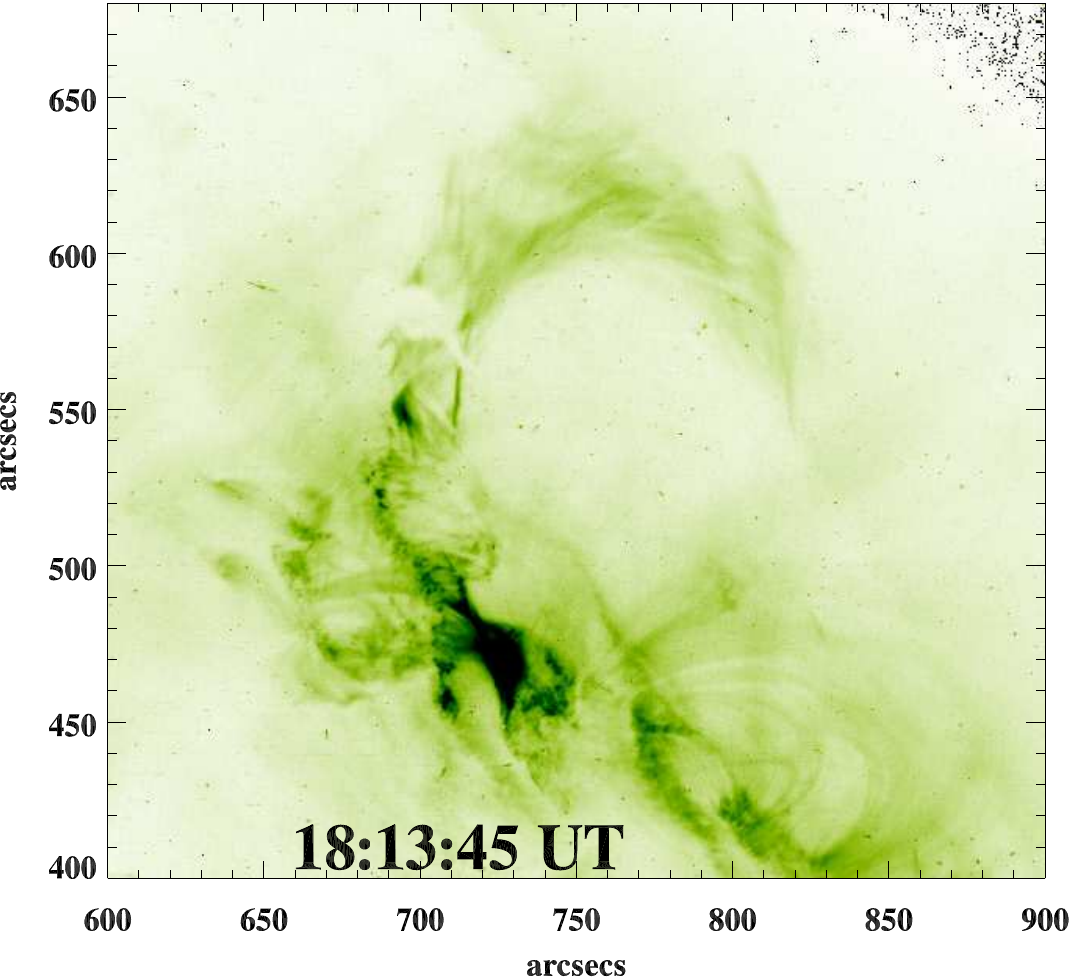}
\thicklines
$ \color{black} \put(-95,75){\vector(0,-2){15}}\put(-100,80){Flare}$
\includegraphics[width=0.5\textwidth]{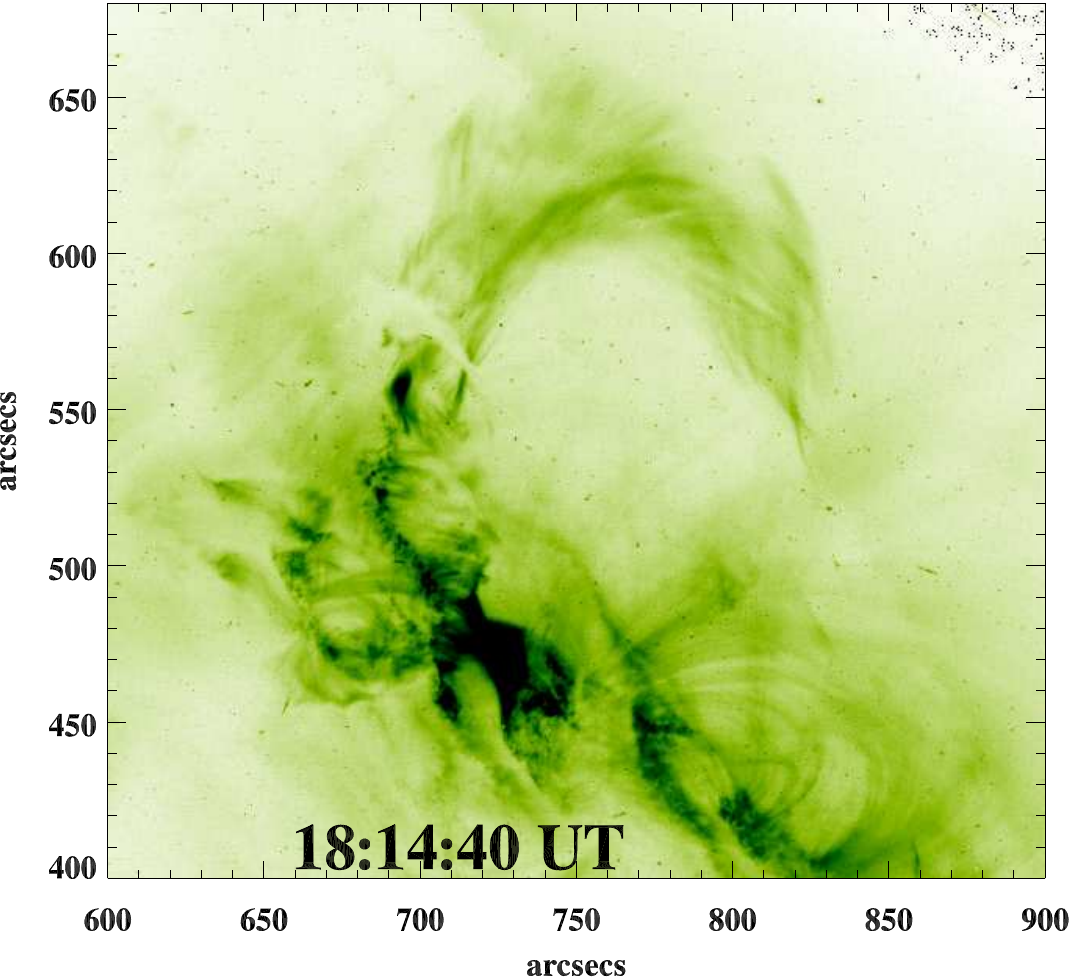}
\thicklines
$ \color{black} \put(-140,110){\vector(1,0){15}}$
\thicklines
$\color{black} \put(-112,108){\circle{23}}$
}
\centerline{
\includegraphics[width=0.5\textwidth]{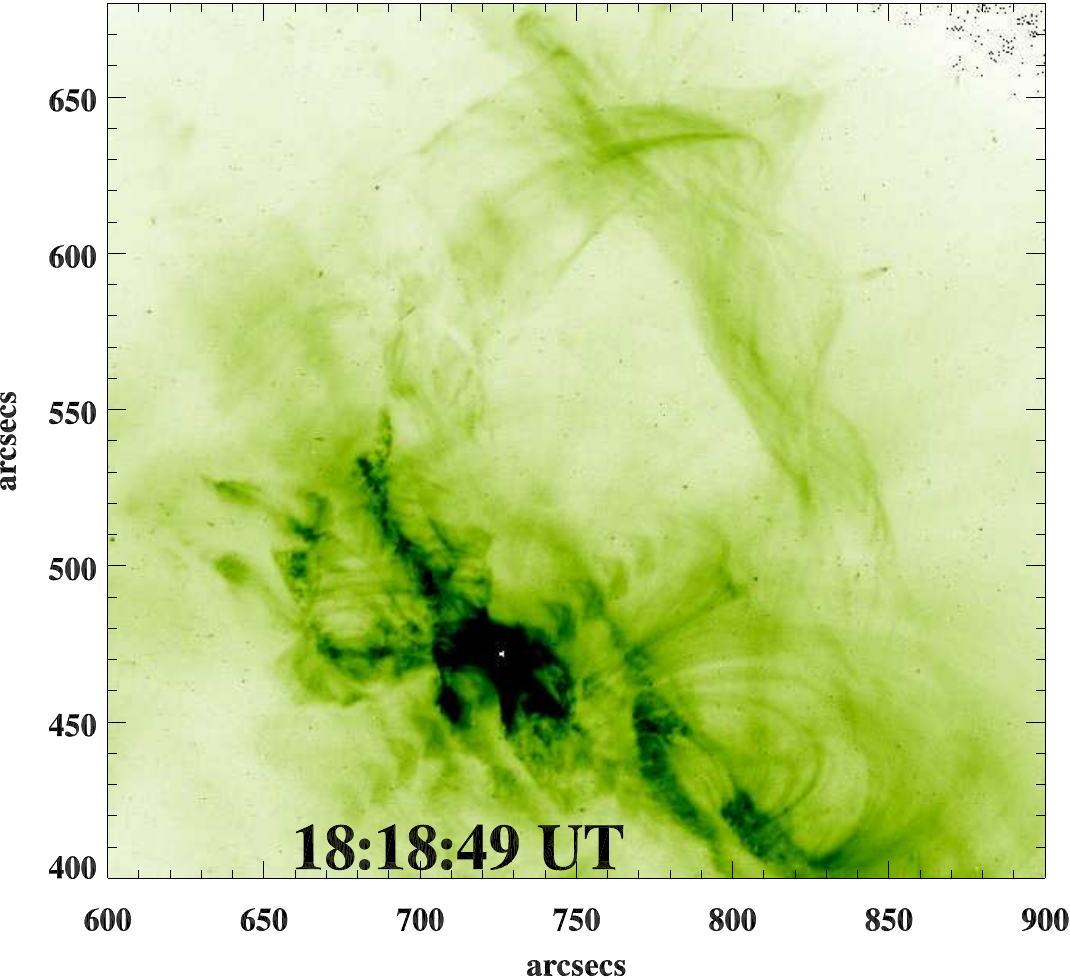}
\includegraphics[width=0.5\textwidth]{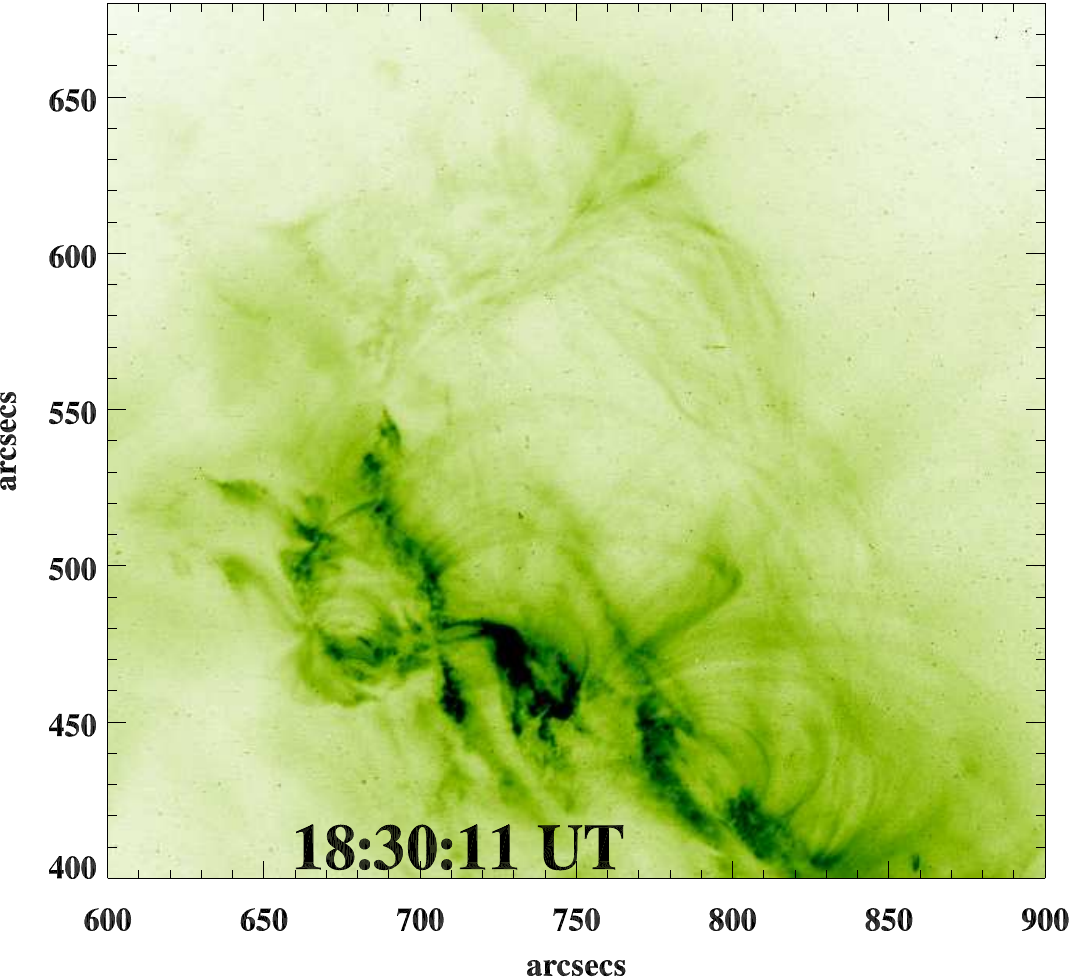}
}
\caption{TRACE 195 \AA \ EUV negative image sequence showing the 
activation of the flux rope, and the triggering of the M1.1/2F flare. The remnant filament is indicated by the arrow in the middle right image.}
\label{tr_195}
\end{figure}

\subsection{TRACE and {\it Hinode}/XRT Observations}
TRACE ({\it Transition Region and Coronal Explorer}) provides the opportunity
to observe the Sun from the chromosphere to the corona \cite{handy1999}.
We have analysed TRACE 195~\AA \ (Fe {\sc xii}, T$\approx$1.5 MK) and
1600~\AA \ ($T$$\approx$4000-10000 K) time sequences.
The field of view for each image is 1024$\times$1024 with 0.5$^{\prime\prime}$ \ pixel$^{-1}$ resolution. 
The  typical cadence for TRACE images is 20-60 s.

Figure \ref{tr_195} displays selected TRACE 195~\AA \ negative images during the flare activity.
The TRACE data have been calibrated and analysed using standard IDL routines in the SolarSoft package. The top-left panel is best interpreted as the potential-field configuration of the active region, with field lines connecting in the form of higher and smaller loop systems. The snapshot at 17:55 UT shows the rising flux rope lying below the eastern small loop-system as indicated by an arrow. The flare was produced near one of the footpoints of the rising flux rope. The thickness of the flux rope in the corona is $\approx$40 Mm. The remnant  filament can also be seen in these images ({\it {\it e.g.}}, at 18:14:40 UT, indicated by the arrow and circle). This filament lay over the rising flux rope. The flux rope showed the twist structure with dips at 18:18:49 UT in the corona. A  supplementary
 TRACE movie shows the suppression of motion in the flux rope, which was possibly associated with the overlying higher loop system in the active region. Then, the plasma filled in the flux rope moved towards the negative polarity where its another footpoint was anchored. The TRACE high resolution observations reveal fine structures of the rising flux tube ({\it e.g.}, the bunch of thin flux tubes showing a kink structure as well as some of nearby loop systems), which were not visible in STEREO SECCHI 171 \AA \ observations. It is only clear from analysing the TRACE images that the rest part of the rope has been repelled and suppressed by the overlying coronal field.

\begin{figure}
\centerline{
\includegraphics[width=0.42\textwidth]{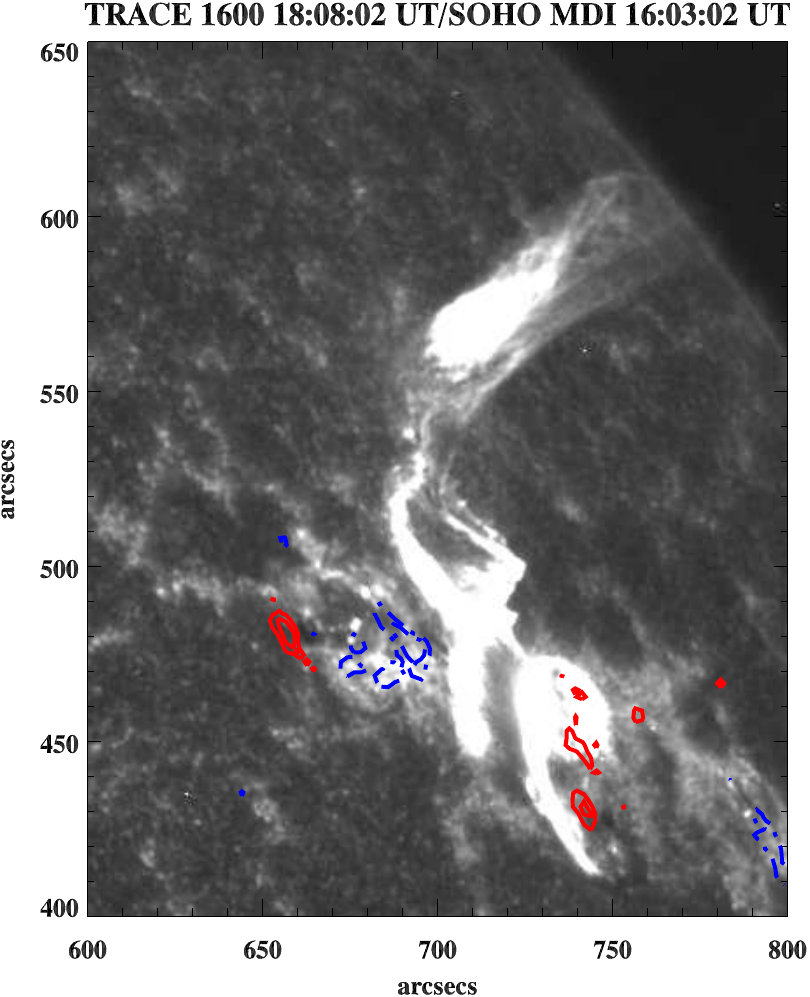}
\thicklines
$ \color{white} \put(-95,100){\vector(1,0){15}}\color{yellow} \put(-130,105){Flux rope}$
$ \color{white} \put(-86,35){\vector(1,1){15}}\color{yellow} \put(-105,30){R1}$
$ \color{white} \put(-35,75){\vector(-1,-1){15}}\color{yellow} \put(-40,80){R2}$
\includegraphics[width=0.6\textwidth]{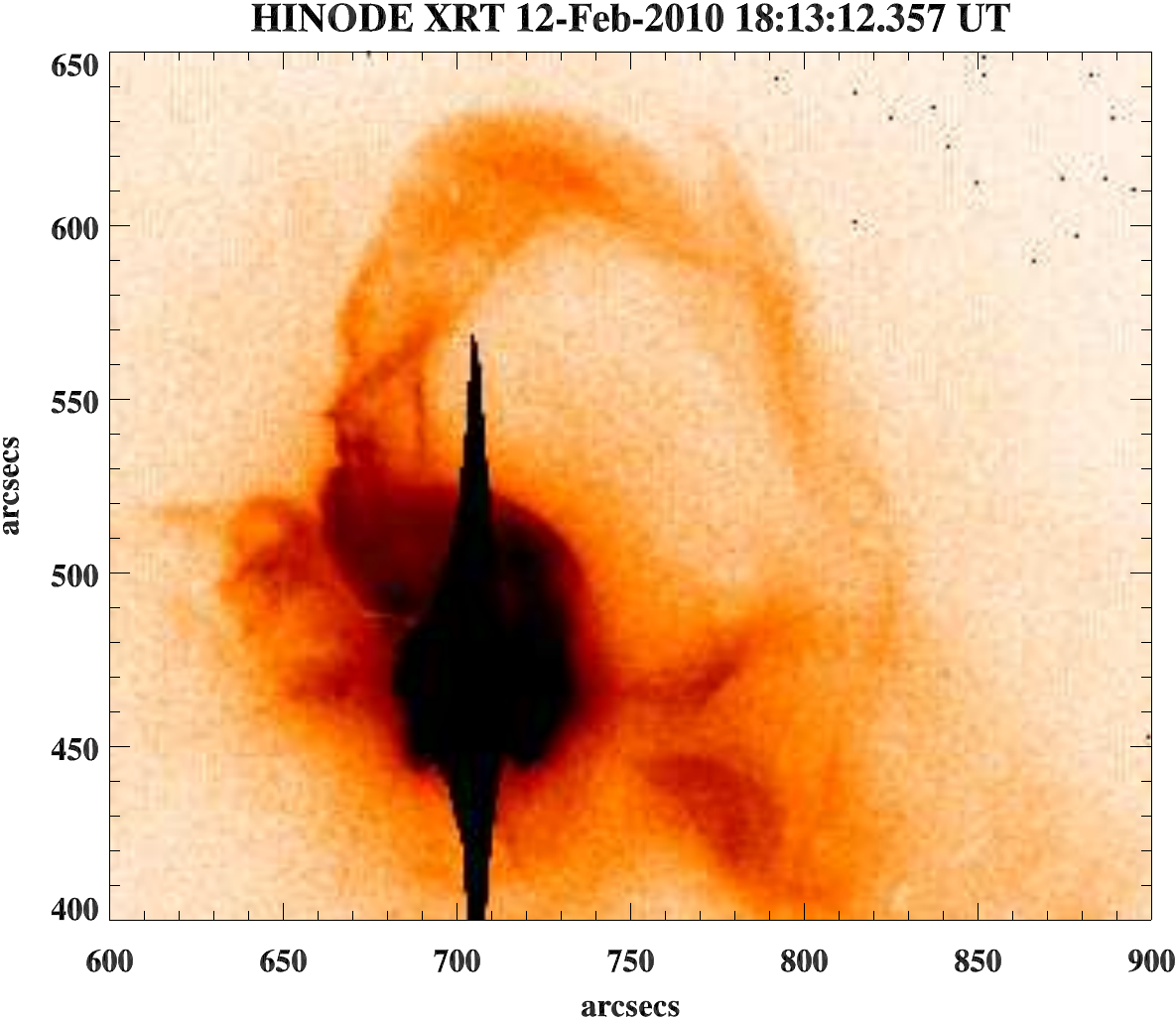}
$ \color{black} \put(-155,35){\vector(1,1){15}}\color{black} \put(-175,28){Flare}$
$ \color{black} \put(-170,120){\vector(1,0){15}}\color{black} \put(-197,125){Flux rope}$
}
\caption{Left: TRACE 1600 \AA \ image overlaid by SOHO/MDI magnetic field contours.
Red and blue show the positive and negative polarities. The flare ribbons `R1' and `R2' are indicated by arrows.
 Right: {\it Hinode} XRT image showing the activation of the flux rope,
 triggering the M1.1/2F flare.}
\label{xrt_tr1600}
\end{figure}
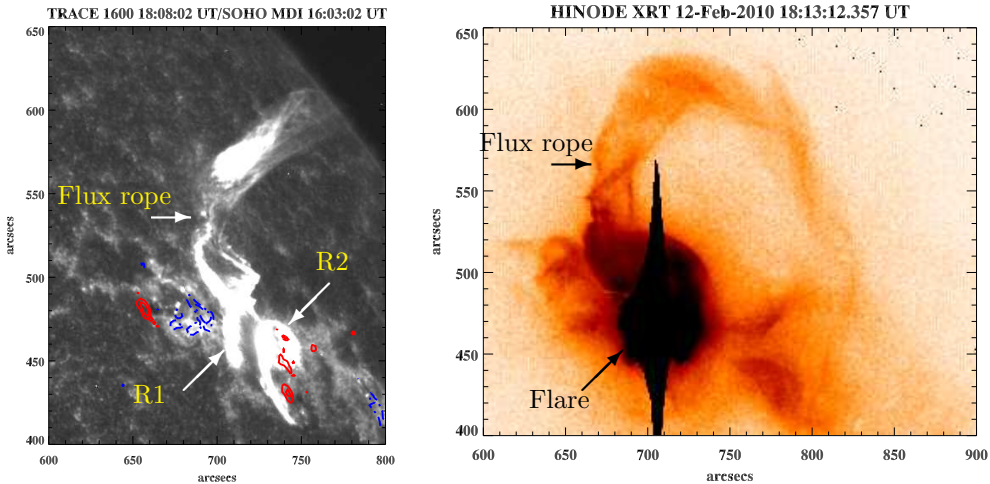

The left panel of Figure \ref{xrt_tr1600} shows TRACE 1600 \AA \ chromospheric data overlaid by SOHO/MDI magnetogram contours (red and blue show positive and negative polarity regions, respectively). The rising bright flux rope is highlighted  by an arrow nearby the negative polarity spot. Two ribbons (indicated by R1 and R2) were formed as the flux rope moved along the PIL. TRACE 1600 \AA \ high resolution observations reveal the two ribbon flare and the activation site of the flux rope near to it. This enables us to understand the spatial configuration of this very unique site where the flux rope activation, the M-class flare and related plasma processes took place. The right panel of Figure \ref{xrt_tr1600} displays a {\it Hinode}/XRT \cite{golub2007} image with the coronal part of the flux rope and the associated flare (indicated by arrows). The activation of magnetic twist is clearly visible in the magnetic--field-dominated soft X-ray corona. The multi-wavelength and simultaneous view of the rising flux rope indicates the multi-temperature plasma filled in it.
\subsection{SOHO/MDI Magnetograms}
In the above subsections we have analysed the rising flux rope from the chromosphere to corona, its suppression due to the remnant filament cavity, and its association with the M-class flare. In the present section, we use SOHO/MDI magnetograms to study the magnetic-field evolution 
before and during the flux rope rising/activation and its associated flare. 
The size of the field of view of each image is 1024$\times$1024 (2$^{\prime\prime}$ pixel resolution) with a cadence of 96 min \cite{sch1995}. We use the standard SolarSoft library to correct for 
the differential rotation and analyse the magnetograms. 

\begin{figure}
\centerline{
\includegraphics[width=0.5\textwidth]{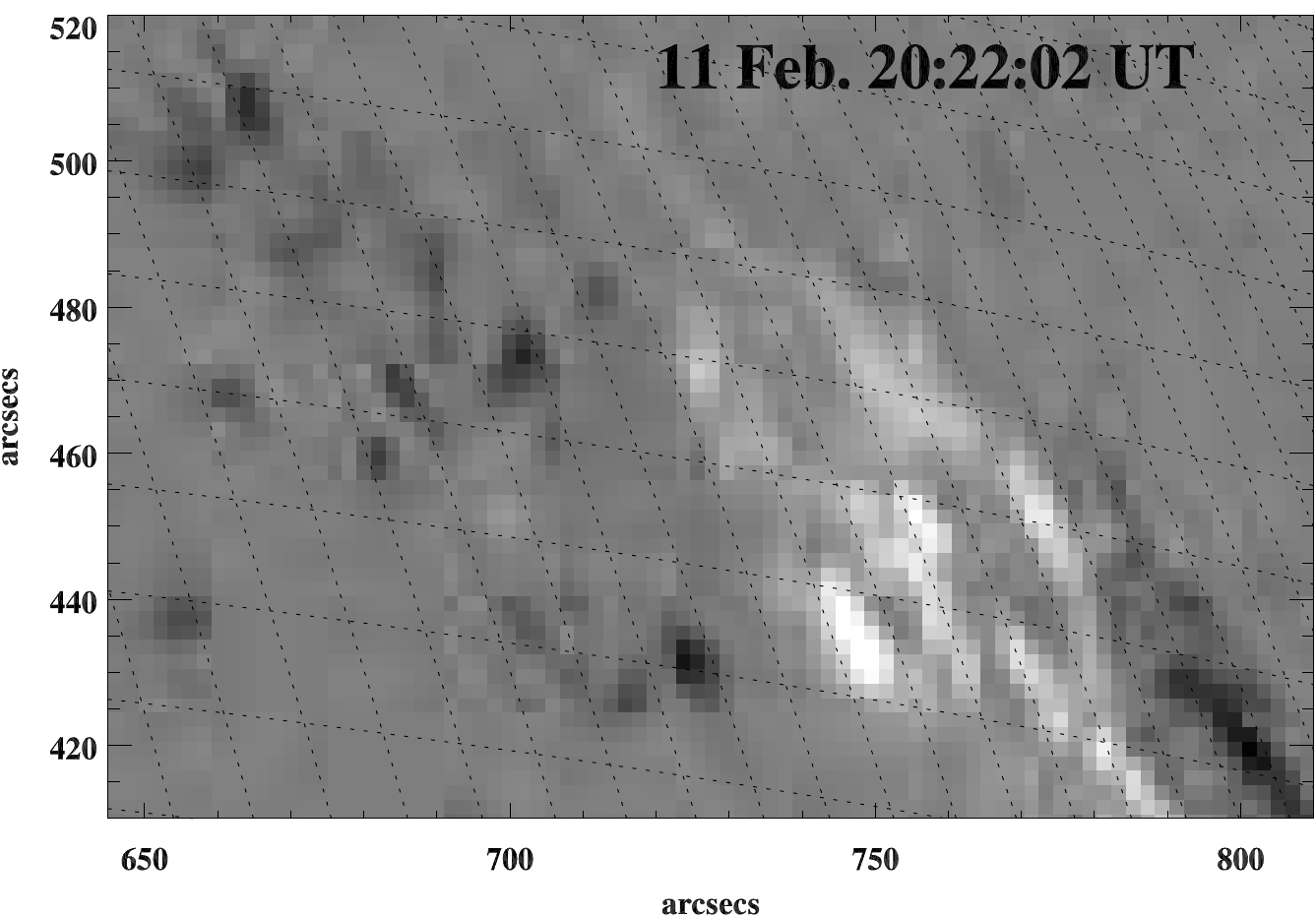}
\includegraphics[width=0.5\textwidth]{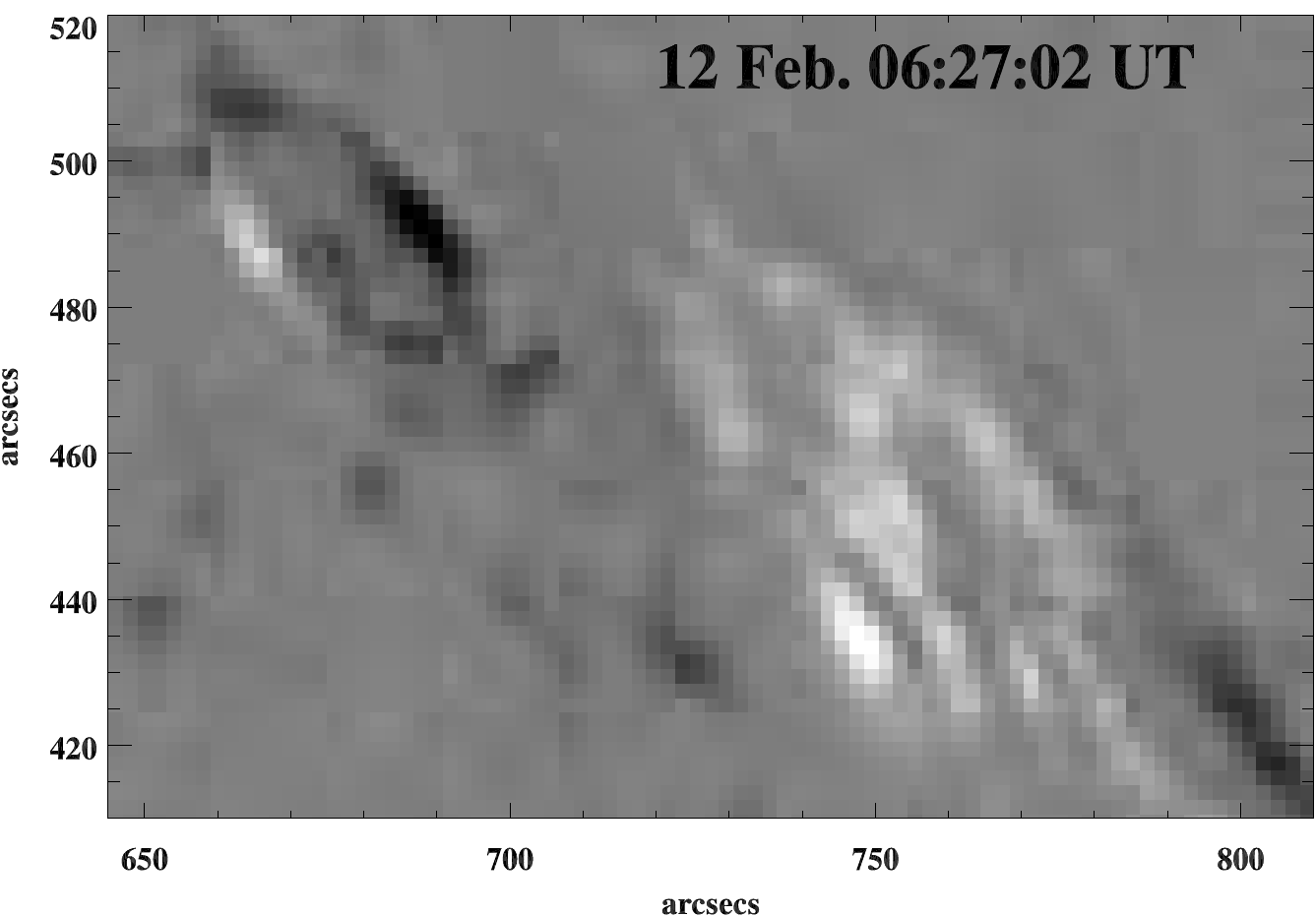}
\thicklines
$\color{yellow} \put(-130,90){\circle{60}}\color{yellow} \put(-160,60){Flux emergence}$
}
\centerline{
\includegraphics[width=0.5\textwidth]{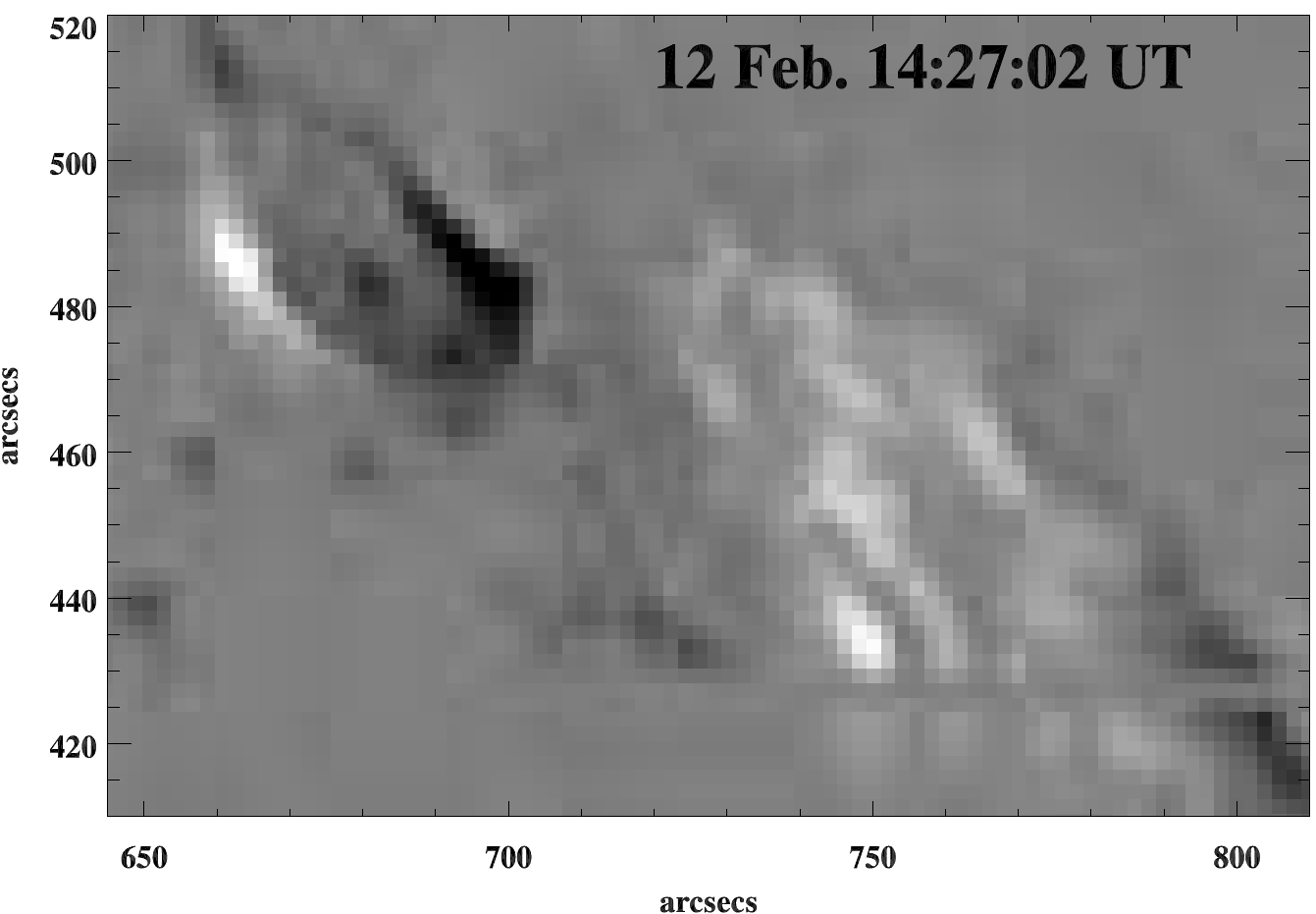}
\thicklines
$ \color{yellow} \put(-160,65){\vector(1,1){12}}$
\includegraphics[width=0.5\textwidth]{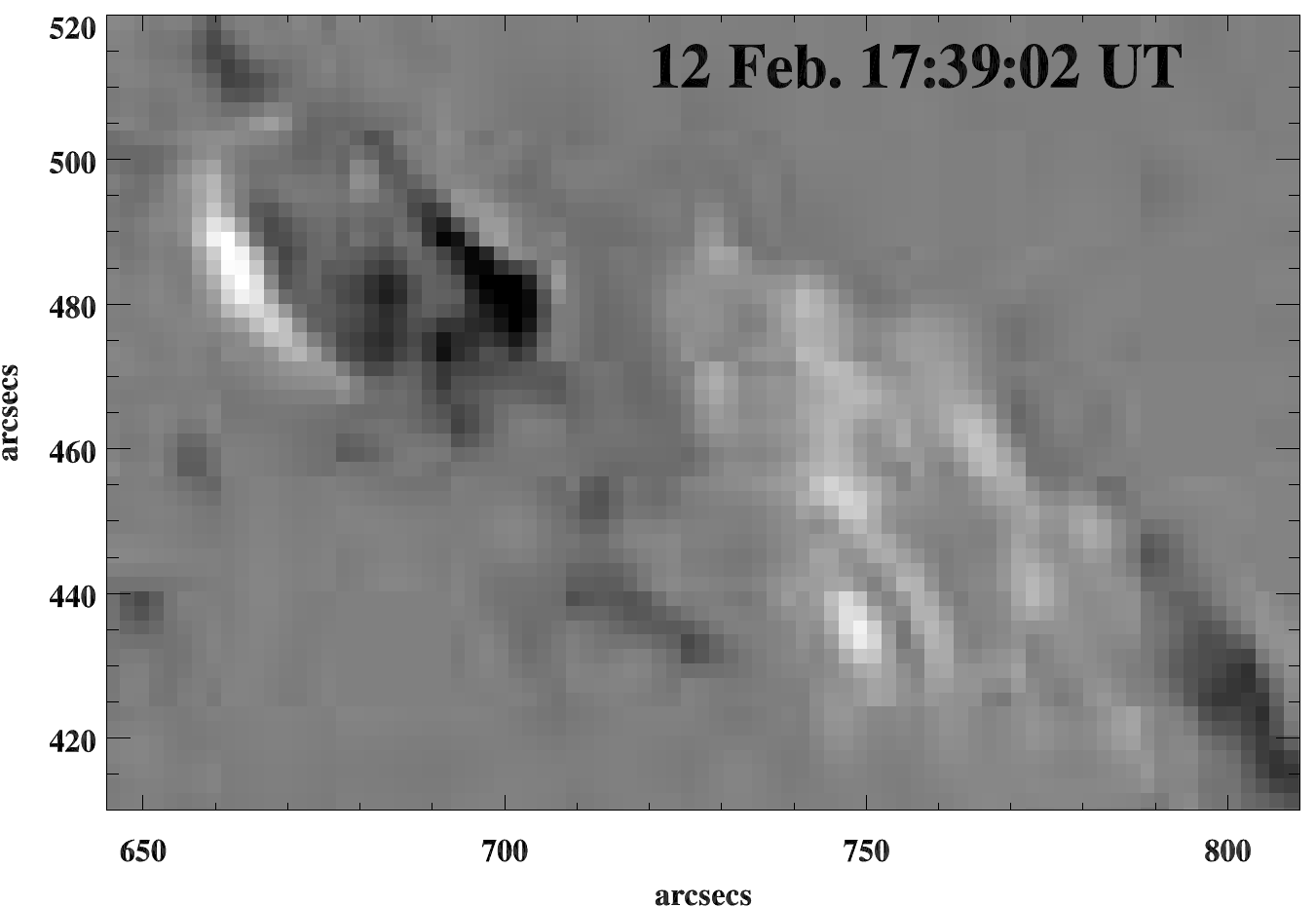}
\thicklines
$ \color{yellow} \put(-160,65){\vector(1,1){12}}$
}
\centerline{
\includegraphics[width=0.5\textwidth]{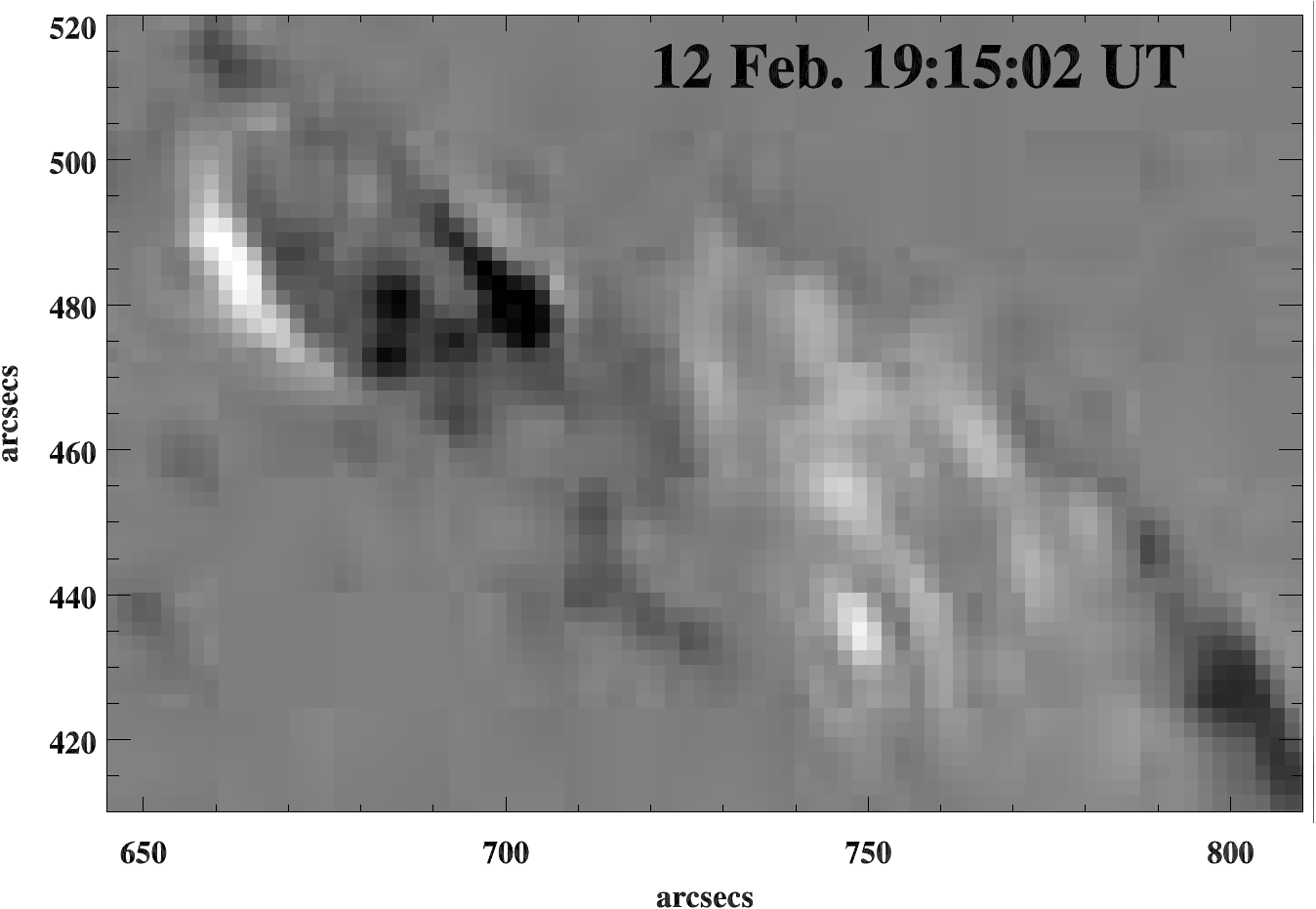}
\thicklines
$ \color{yellow} \put(-160,65){\vector(1,1){12}}$
\thicklines
$\color{yellow} \put(-125,75){\circle{23}}$
\includegraphics[width=0.5\textwidth]{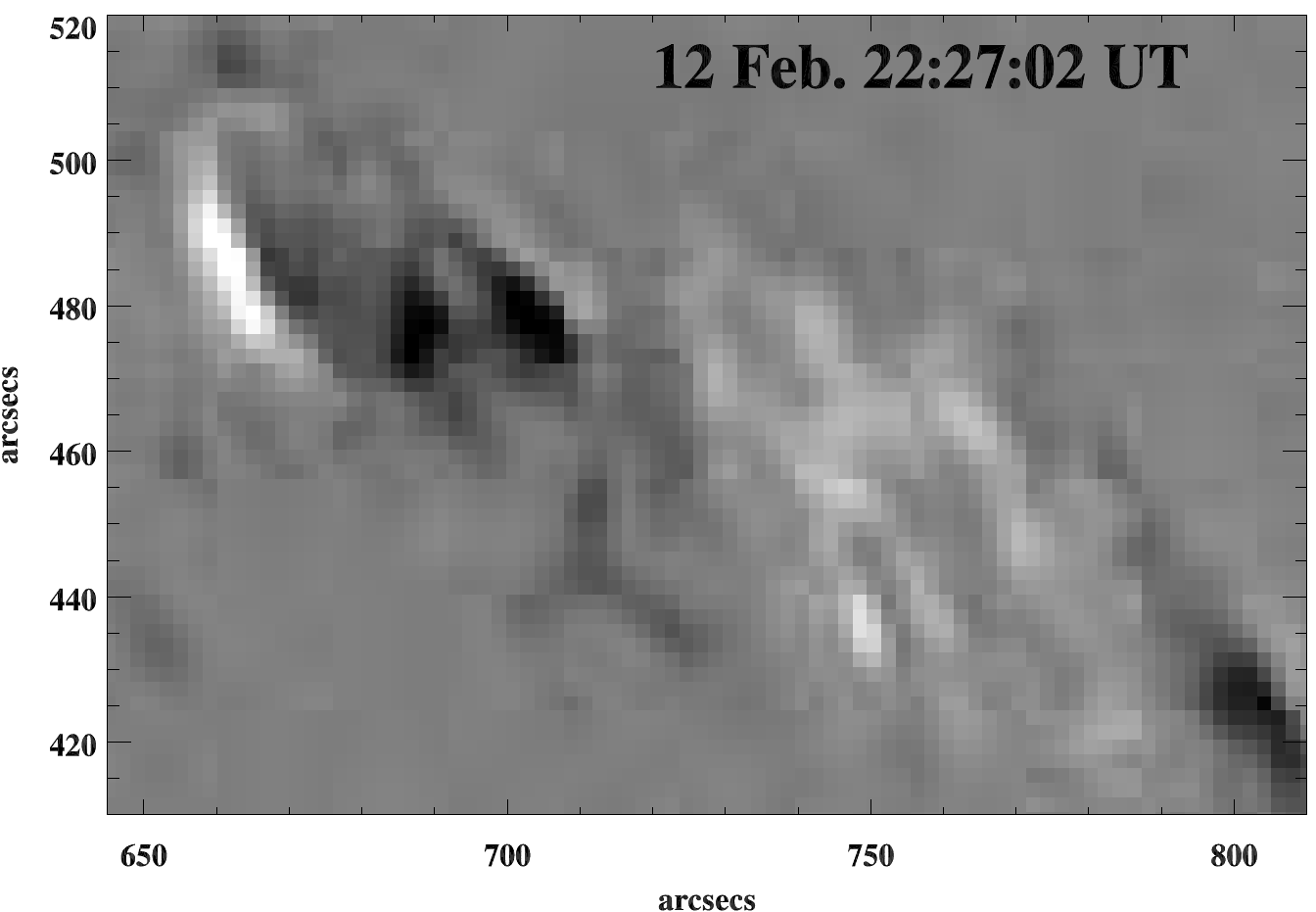}
\thicklines
$ \color{yellow} \put(-160,65){\vector(1,1){12}}$
$\color{yellow} \put(-125,75){\circle{23}}$
}
\caption{SOHO/MDI magnetograms of the flux rope emergence location. The top-right image shows the flux emergence of opposite magnetic polarities (indicated by circles). Arrows indicate the positive polarity flux emergence. The images in the bottom panel show the emergence of negative polarity region (indicated by circles).}
\label{mdi}
\end{figure}

Figure \ref{mdi} 
displays the sequence of MDI images on 11 and 12 February 2010. The MDI 
time-sequence reveals some very interesting features of the flux emergence. The top-right image clearly
shows the emergence of two opposite polarities (indicated inside the circles) 
on 12 February 2010. This newly emerged flux grew continuously later on. Next, a 
considerable change in the emerging 
polarities (indicated by arrow) is demonstrated. The negative polarity showed motion in 
the south-west direction, whereas the positive polarity migrated in the north-east direction.
The flux rope structure rose in the western part of the growing negative polarity region.
The flux rope was activated and moved up in the eastern side of a widely spread positive polarity 
magnetic flux concentrations near the PIL.
The motion of the negative polarity towards the flare and flux rope activation site is 
also observed in the MDI time sequence.

\begin{figure} 
\centerline{ 
\includegraphics[width=0.5\textwidth]{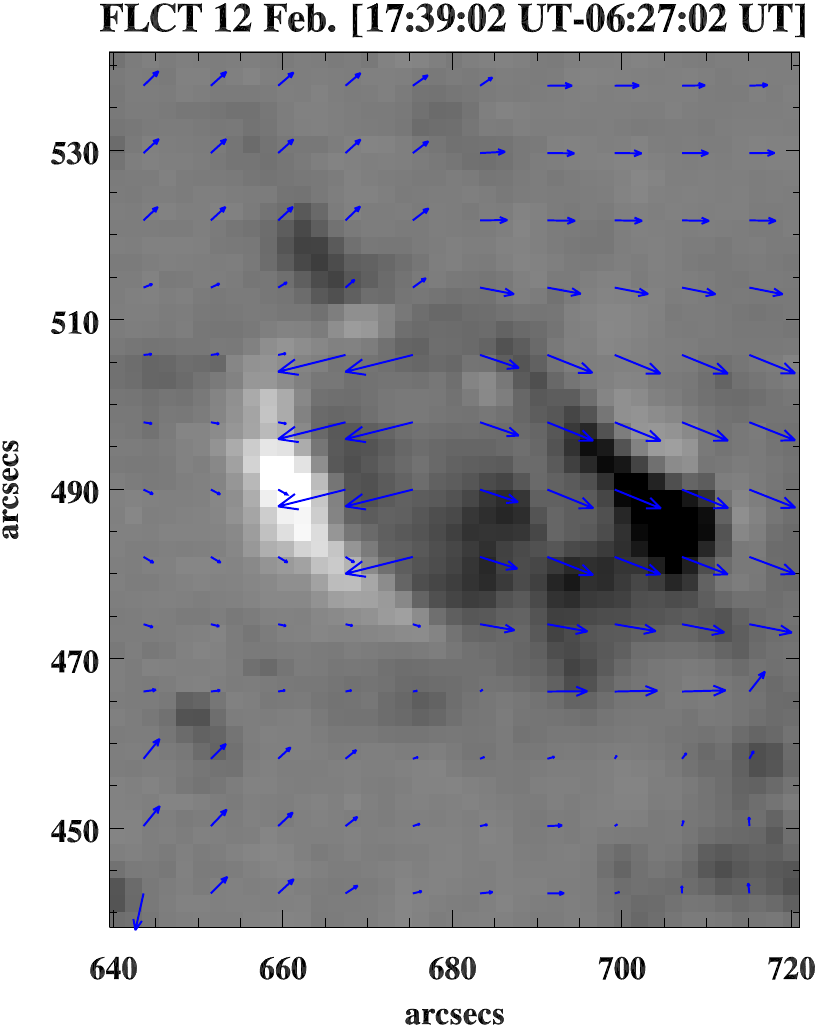} 
} 
\caption{FLCT photospheric flow map derived from MDI magnetograms on 12 February 2010. The longest arrow corresponds to the flow speed of $\approx$335 m s$^{-1}$.} 
\label{flct} 
\end{figure} 

To confirm this motion that is very crucial to understand flare energy build-up, we measure the photospheric
horizontal flow pattern in and around the emerging bipole in the active region. We use
the Fourier Local Correlation Tracking Technique (FLCT) on
SOHO/MDI data sequences for this purpose \cite{welsch2007,fisher2008}. The main input parameters for this technique
are two images, f1 and f2, the pixel separation ($\Delta$s),
time separation ($\Delta$t), and a Gaussian window size
($\sigma$). This routine calculates the (two dimensional) velocity by
maximizing the cross-correlation between the two images when weighted
by a Gaussian window centered on each pixel location. In
our study, we use the two SOHO/MDI frames at different times
before the flare. After a careful investigation, a Gaussian window with $\sigma=15^{\prime\prime}$ was chosen. Figure \ref{flct} displays the photospheric velocity map obtained
by the FLCT technique using SOHO/MDI magnetograms. The
longest arrow corresponds to a velocity of 0.335 km s$^{-1}$. It may
be noted that the map shows the opposite flows in the emerging bipole.
The negative polarity moved in the south-west direction where the flux rope activation and flare occurred, whereas the positive polarity  
move oppositely in the eastward direction. 
This motion may develop twist in the emerging bipole and store the magnetic
energy for the event.  As we have seen, the energy was finally released due to the flux rope activation and its reconnection with the
pre-existing magnetic field.

\section{Magnetic Configuration and Scenario of the Event}
\begin{figure}
\centerline{
\includegraphics[width=0.55\textwidth]{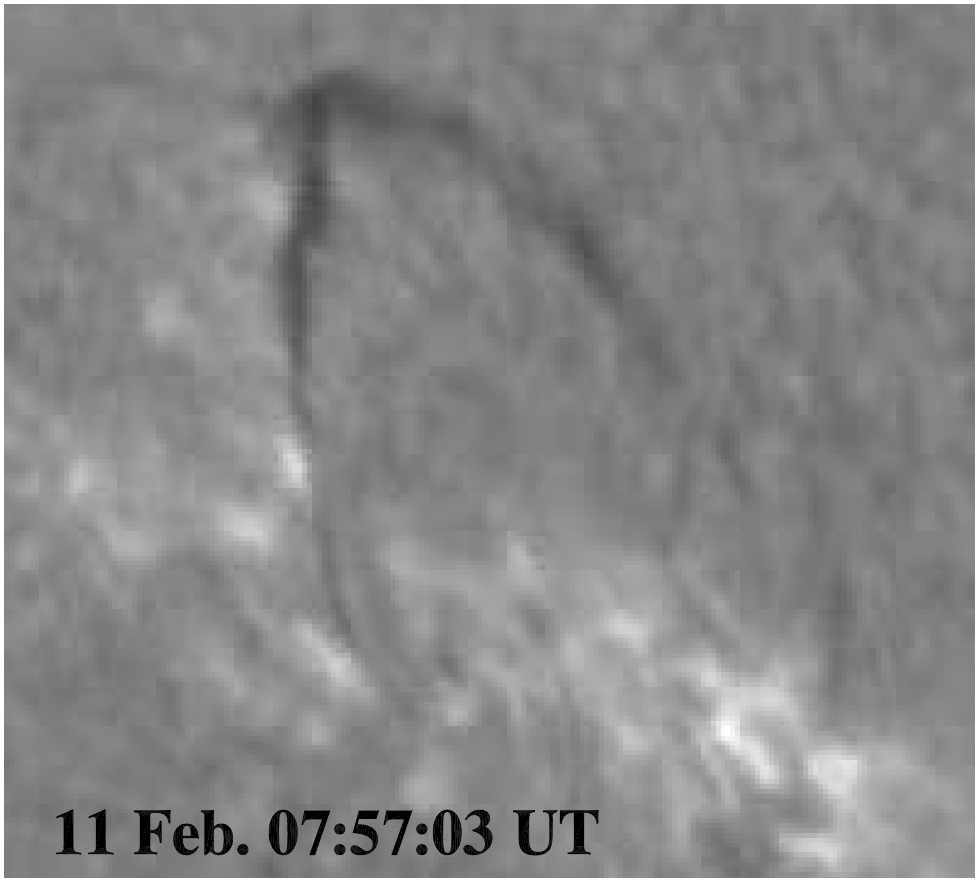}
}
\caption{Semi-circular filament in the active region NOAA 11045 on 11 February 2010. (Courtesy of Catania Astrophysical Observatory). The size of the image is 200$^{\prime\prime}$$\times$180$^{\prime\prime}$.}
\label{ha}
\end{figure}
In Section 2, we described the multi-wavelength and magnetic field observations of the rising twisted flux rope 
and the associated M-class flare. In the present section, we elaborate the magnetic field configuration of the flaring region 
and related processes in the light of the multi-wavelength observational scenario.

 A semi-circular filament was observed on the northern border of the active region NOAA 11045 couple of days before the event (Figure \ref{ha}). It had been located at the same place where the twisted EUV structure and the bright H$\alpha$ arch were observed on 12 February 2010, and the shape was also the same. On the previous days the eastern side of the filament was clearly related to a positive polarity sunspot. Some connectivity is still visible in Figure \ref{ha}, particularly on 11 February 2010. Comparison with the BBSO filtergrams on 12 February, before the event, shows that only the eastern part of the filament existed on 11 February 2010. However, propagation of brightening along a trajectory that coincided with the shape of the whole semi-circular filament suggests that the magnetic structure remained nearly the same during these two days despite the absence of visible plasma in the western part of the structure.

\begin{figure}
\centerline{
\includegraphics[width=0.7\textwidth]{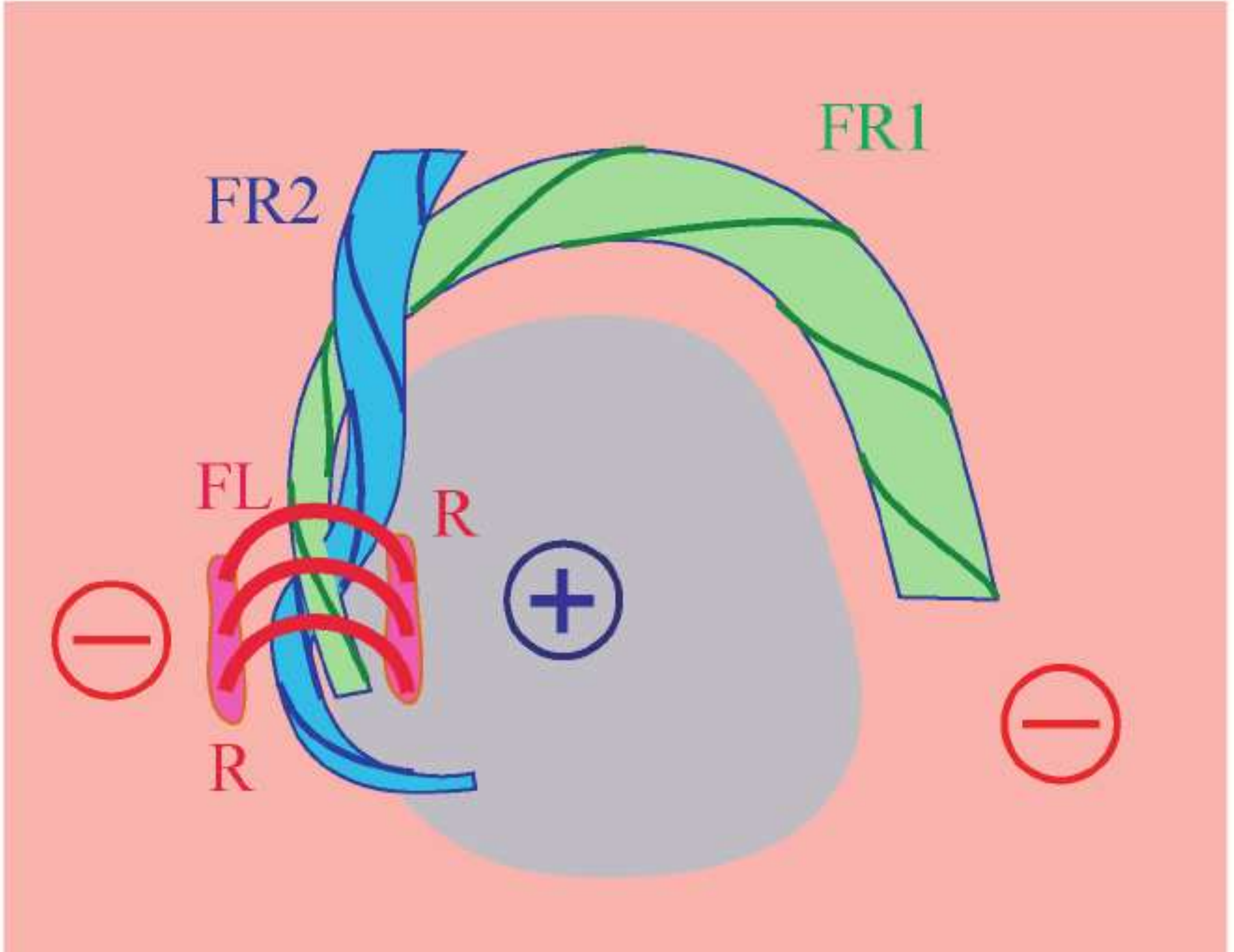}
}
\caption{Schematic representation of the magnetic configuration of the active region NOAA 11045. `FR1' and `FR2' are parts of the flux rope containing the filament. `R' and `FL' represent the flare ribbons and flare loops respectively.}
\label{cartoon}
\end{figure}

We may assume that a twisted magnetic flux rope existed in the northern area of the active region NOAA 11045 (see a cartoon outlined in Figure \ref{cartoon}). One end of the flux rope was rooted near the positive polarity sunspot, while the other end was rooted in the area of negative polarity westward of the area of positive polarity. The flux rope was aligned along the polarity inversion line and surrounding the positive polarity on the north. On 11 February, the whole length of the flux rope became filled with cold and dense plasma, which constituted the filament. On the next day, only the eastern part of the flux rope was filled with plasma and it looked as if a separate strand of the flux rope (marked as FR2 in Figure \ref{cartoon}) was rooted at some place to the north of the polarity inversion line near the middle of the flux rope.

At 17:44 UT on 12 February, activation of the flux rope started near its eastern end, which manifested itself by motion and heating of the filament material. There is no clear evidence of rising motion of the whole body of this filament segment. Plasma heated to different temperatures spread along the field lines of the flux rope. The multi-temperature nature of the event allows one to see the process in different wavelengths from H$\alpha$ to soft X-rays. The activation affected only those strands of the flux rope (marked as FR1 in the Figure \ref{cartoon}) that were connected to the area of negative polarity to the west of the positive polarity. Thus, we can assume that the two bunches of strand were individual flux ropes, FR1 and FR2, braided together. The northern section of FR2 passed over the FR1 and stayed undisturbed during the propagation of brightening along the FR1 below it. The plasma in FR2 plasma was seen as a dark filament in H$\alpha$ and as a dark loop-like feature crossing bright FR1 threads in EUV images. Since the northern end of FR2 was rooted close to the place of crossing of the two ropes, possibly, this anchoring have prevented FR1 from rising up and erupting. At a late phase of the event, FR2 became also activated. It was slightly shifted up and became heated. This is the reason why this arch disappeared in H$\alpha$ and EUV images in the middle of the event but appeared again after its ending. 

The bright twisted threads in FR1, after coming out from under FR2, expanded into a wide structure that at first increases its height but then decended. Plasma flowed along helical flux tubes to the western end of the flux rope, while the flux tubes themselves showed complex untwisting and writhing motion.

After 18:00 UT, two short flare ribbons began to progress near the site of the initial filament activation. Soon bright flare loops, connecting the ribbons, appeared above the filament. This is not typical for two-ribbon solar flares. Usually flare loops appear below erupting filaments. However, there are other examples of such a behavior. For example, flare loops were seen above the stable part of an eruptive filament in the complex event on 20 September 1998 (see TRACE 171 \AA \ movie $\#$ 21 at the TRACE site http://trace.lmsal.com/POD/). The flux rope images of STEREO A/SECCHI in 171 \AA \ and 304 \AA \ ({\it cf.}, last snapshots of Figures 3 and 5) validate the schematic diagram and magnetic field environment.

The process of the initial filament activation is of smaller spatial scale and its mechanism is yet unclear to us. However, we suggest that rapid changes in the magnetic field close to this place are the main factor that influences the initiation of the filament activation and the flare. Rapid emergence of new magnetic flux and fast motion of the negative polarity in the direction to the opposite polarity towards PIL squeezed significantly the flux rope and must have lead to some instability and energy release.

The event might be called a failed filament eruption but the absence of filament material in the significant portion of the flux rope before the activation makes this appellation ambiguous. It is better to refer to this event as a failed flux rope eruption. The ambient magnetic field may have been strong enough to prevent the flux rope from erupting. The flux rope was presumably in a stable equilibrium. The disturbance have led only to oscillations near the equilibrium position as well as untwisting and writhing motion.

\section{Discussion and Conclusions}
In the present paper, we showed a rare multi-wavelength observation of a rising twisted flux rope on 12 February 2010 from NOAA AR 11045. The flux rope was activated in the vicinity of 
a remnant filament structure and moved up in the corona as 
visible in H$\alpha$, EUV and soft X-ray observations. An M-class flare was triggered as the leading edge of the brightening moved away from the active region 
with a (projected) speed of $\approx$90 km s$^{-1}$.  A two-ribbon flare progressed
well with the ascent of the flux rope.  Most likely, the lower part of the 
flux rope structure reconnected with the surrounding low-lying loop system causing 
the flare. The flux rope reached to the maximum height and then 
it was suppressed by the overlying fields as the bending of the rope was observed in TRACE EUV observations. 
An analysis of associated MDI magnetograms show the emergence of a dipole $\approx$12 h prior to the flare and flux rope activation. 
The continuous growth and motion of two spots in the opposite direction (particularly the negative polarity spot), helped 
in building the magnetic energy necessary to initiate a flare in the vicinity of the negative spot, where the flux rope activation 
took place. 
The structure moved up near the polarity inversion line and its end was anchored to the negative polarity field region resulting failed eruption.
 
Using SOT (Solar Optical Telescope) onboard {\it Hinode}, vector magnetic 
field observations of NOAA AR 10953, \inlinecite{okamoto2008} found the 
signature of an emerging helical flux rope beneath an active region prominence. 
They have found that a helical flux rope was emerging from below the 
photosphere into the corona along the PIL under the pre-existing prominence, and, suggested that the emergence of the helical magnetic flux rope was 
associated with the evolution and maintenance of the prominence as after 
the emergence of the flux rope prominence became more stable. In our 
case, we do not see any evidence of positive flux emergence near 
the flare and flux rope activation site. However, we have found such signature 
on the eastern side of PIL. Instead, a unique situation was created in the form 
of the activation of twisted flux rope that was going below the remnant 
filament channel and being failed finally. Further, the overlying filament 
had not changed much during the rise of the flux rope. The reconnection of this rope with the surrounding field probably led to 
the flare energy release at the activation site. The rise of the helical 
flux rope that underwent reconnection with lower coronal fields, had possibly 
carried material into the corona and remnant filament cavity as recently 
observed \cite{okamoto2009,okamoto2010}. The rising motion 
of the flux rope ($\approx$90 km s$^{-1}$) was possibly related to the 
outflow driven by the magnetic reconnection between the flux rope and 
the pre-existing coronal field in AR 11045 at the flare site.

In conclusion, we have presented here the first multi-wavelength observation of the unique event of a failed flux rope most likely due to the remnant filament in the beginning and then by an overlying coronal magnetic field. This interesting dynamics in the vicinity of rapidly emerging bipolar region and motion of its negative polarity towards the flux tube activation site may have provided the initial conditions for the flare energy build-up, and then its release due to the reconnection of the rising flux tube with the pre-existing field lines at the flare site. Further and more elaborative multi-wavelength campaign should  be carried out using the latest space ({\it e.g.}, {\it Hinode}, STEREO, SDO \cite{graham2003,lemen2011}) and ground-based ({\it e.g.}, SST, ROSA \cite{scharmer2003,jess2010}) observatories to shed more light on such unique energy build up and release processes of solar flares associated with rapidly emerging active regions.
 
\begin{acks}
We thank the reviewer for his/her suggestions for our present research work. We acknowledge space missions, GOES, SOHO/MDI, {\it Hinode}/XRT, TRACE and STEREO for providing the data used in this study.  SOHO is a project of international cooperation between ESA and NASA. {\it Hinode} is a Japanese mission developed and launched by
ISAS/JAXA, collaborating with NAOJ as a domestic partner, NASA and STFC (UK) as
international partners.
  We are thankful to BBSO and Catania Astrophysical Observatory for
providing the H$\alpha$ data used in this study. Global High Resolution
H$\alpha$ Network is operated by the Space Weather Research Lab, New
Jersey Institute of Technology. This work was supported by the Department of
Science and Technology, Ministry of Science and Technology of India,
by the Russian foundation for Basic Research (grants 09-02-00080 and
09-02-92626, INT/RFBR/P-38). RE acknowledges M. Ker\'ay for patient encouragement and is also grateful
to NSF, Hungary (OTKA, Ref. No. 483133) for financial support received. AKS thanks SP2RC, School of Mathematics and Statistics, The Univesity of Sheffield, U.K. for
the support of collaborative visits, where
the part of present research work has been carried out.
 
\end{acks}
\bibliographystyle{spr-mp-sola}
\bibliography{reference}  

\begin{thebibliography}{46}
\ifx \bisbn   \undefined \def \bisbn  #1{ISBN #1}\fi
\ifx \binits  \undefined \def \binits#1{#1}\fi
\ifx \bauthor  \undefined \def \bauthor#1{#1}\fi
\ifx \batitle  \undefined \def \batitle#1{#1}\fi
\ifx \bjtitle  \undefined \def \bjtitle#1{\textit{#1}}\fi
\ifx \bvolume  \undefined \def \bvolume#1{\textbf{#1}}\fi
\ifx \byear  \undefined \def \byear#1{#1}\fi
\ifx \bissue  \undefined \def \bissue#1{#1}\fi
\ifx \bfpage  \undefined \def \bfpage#1{#1}\fi
\ifx \blpage  \undefined \def \blpage #1{#1}\fi
\ifx \burl  \undefined \def \burl#1{\textsf{#1}}\fi
\ifx \href  \undefined \def \href#1#2{\textsf{#2}}\fi
\ifx \doiurl  \undefined \def
  \doiurl#1{\href{http://dx.doi.org/#1}{\textsf{#1}}}\fi
\ifx \betal  \undefined \def \betal{\textit{et al.}}\fi
\ifx \binstitute  \undefined \def \binstitute#1{#1}\fi
\ifx \bctitle  \undefined \def \bctitle#1{#1}\fi
\ifx \beditor  \undefined \def \beditor#1{#1}\fi
\ifx \bpublisher  \undefined \def \bpublisher#1{#1}\fi
\ifx \bbtitle  \undefined \def \bbtitle#1{\textit{#1}}\fi
\ifx \bedition  \undefined \def \bedition#1{#1}\fi
\ifx \bseriesno  \undefined \def \bseriesno#1{\textbf{#1}}\fi
\ifx \blocation  \undefined \def \blocation#1{#1}\fi
\ifx \bsertitle  \undefined \def \bsertitle#1{\textit{#1}}\fi
\ifx \bsnm \undefined \def \bsnm#1{#1}\fi
\ifx \bsuffix \undefined \def \bsuffix#1{#1}\fi
\ifx \bparticle \undefined \def \bparticle#1{#1}\fi
\ifx \barticle \undefined \def \barticle{}\fi
\ifx \botherref \undefined \def \botherref{}\fi
\ifx \url \undefined \def \url#1{\textsf{#1}}\fi
\ifx \bchapter \undefined \def \bchapter{}\fi
\ifx \bbook \undefined \def \bbook{}\fi
\ifx \bcomment \undefined \def \bcomment#1{#1}\fi
\ifx \oauthor \undefined \def \oauthor#1{#1}\fi
\ifx \citeauthoryear \undefined \def \citeauthoryear#1{#1}\fi
\def \endbibitem {}

\bibitem[\protect\citeauthoryear{{Alexander}, {Liu}, and
  {Gilbert}}{2006}]{alex2006}
\begin{barticle}
\bauthor{\bsnm{{Alexander}}, \binits{D.}}, \bauthor{\bsnm{{Liu}}, \binits{R.}},
  \bauthor{\bsnm{{Gilbert}}, \binits{H.R.}}:
\byear{2006},
\batitle{{Hard X-Ray production in a failed filament eruption}}.
\bjtitle{\apj}
\bvolume{653},
\bfpage{719}\,--\,\blpage{724}.
doi:\doiurl{10.1086/508137}.
\end{barticle}
\endbibitem

\bibitem[\protect\citeauthoryear{{Amari} \textit{et~al.}}{2000}]{amari2000}
\begin{barticle}
\bauthor{\bsnm{{Amari}}, \binits{T.}}, \bauthor{\bsnm{{Luciani}},
  \binits{J.F.}}, \bauthor{\bsnm{{Mikic}}, \binits{Z.}},
  \bauthor{\bsnm{{Linker}}, \binits{J.}}:
\byear{2000},
\batitle{{A twisted flux rope model for coronal mass ejections and two-ribbon
  flares}}.
\bjtitle{Astrophys. J. Lett.}
\bvolume{529},
\bfpage{49}\,--\,\blpage{52}.
doi:\doiurl{10.1086/312444}.
\end{barticle}
\endbibitem

\bibitem[\protect\citeauthoryear{{Archontis}, {Hood}, and
  {Brady}}{2007}]{arch2007}
\begin{barticle}
\bauthor{\bsnm{{Archontis}}, \binits{V.}}, \bauthor{\bsnm{{Hood}},
  \binits{A.W.}}, \bauthor{\bsnm{{Brady}}, \binits{C.}}:
\byear{2007},
\batitle{{Emergence and interaction of twisted flux tubes in the Sun}}.
\bjtitle{\aap}
\bvolume{466},
\bfpage{367}\,--\,\blpage{376}.
doi:\doiurl{10.1051/0004-6361:20066508}.
\end{barticle}
\endbibitem

\bibitem[\protect\citeauthoryear{{Archontis} \textit{et~al.}}{2004}]{arch2004}
\begin{barticle}
\bauthor{\bsnm{{Archontis}}, \binits{V.}}, \bauthor{\bsnm{{Moreno-Insertis}},
  \binits{F.}}, \bauthor{\bsnm{{Galsgaard}}, \binits{K.}},
  \bauthor{\bsnm{{Hood}}, \binits{A.}}, \bauthor{\bsnm{{O'Shea}}, \binits{E.}}:
\byear{2004},
\batitle{{Emergence of magnetic flux from the convection zone into the
  corona}}.
\bjtitle{\aap}
\bvolume{426},
\bfpage{1047}\,--\,\blpage{1063}.
doi:\doiurl{10.1051/0004-6361:20035934}.
\end{barticle}
\endbibitem

\bibitem[\protect\citeauthoryear{{Archontis} \textit{et~al.}}{2005}]{arch2005}
\begin{barticle}
\bauthor{\bsnm{{Archontis}}, \binits{V.}}, \bauthor{\bsnm{{Moreno-Insertis}},
  \binits{F.}}, \bauthor{\bsnm{{Galsgaard}}, \binits{K.}},
  \bauthor{\bsnm{{Hood}}, \binits{A.W.}}:
\byear{2005},
\batitle{{The three-dimensional interaction between emerging magnetic flux and
  a large-scale coronal field: reconnection, current Sheets, and jets}}.
\bjtitle{\apj}
\bvolume{635},
\bfpage{1299}\,--\,\blpage{1318}.
doi:\doiurl{10.1086/497533}.
\end{barticle}
\endbibitem

\bibitem[\protect\citeauthoryear{{Aschwanden}}{2004}]{asc2004}
\begin{bbook}
\bauthor{\bsnm{{Aschwanden}}, \binits{M.J.}}:
\byear{2004},
\bbtitle{{Physics of the Solar Corona. An Introduction}},
\bpublisher{Praxis Publishing Ltd, Springer}, \blocation{Berlin}.
\end{bbook}
\endbibitem

\bibitem[\protect\citeauthoryear{{Bonet} \textit{et~al.}}{2008}]{bonet2008}
\begin{barticle}
\bauthor{\bsnm{{Bonet}}, \binits{J.A.}}, \bauthor{\bsnm{{M{\'a}rquez}},
  \binits{I.}}, \bauthor{\bsnm{{S{\'a}nchez Almeida}}, \binits{J.}},
  \bauthor{\bsnm{{Cabello}}, \binits{I.}}, \bauthor{\bsnm{{Domingo}},
  \binits{V.}}:
\byear{2008},
\batitle{{Convectively driven vortex flows in the Sun}}.
\bjtitle{Astrophys. J. Lett.}
\bvolume{687},
\bfpage{131}\,--\,\blpage{134}.
doi:\doiurl{10.1086/593329}.
\end{barticle}
\endbibitem

\bibitem[\protect\citeauthoryear{{Fan}}{2001}]{fan2001}
\begin{barticle}
\bauthor{\bsnm{{Fan}}, \binits{Y.}}:
\byear{2001},
\batitle{{The emergence of a twisted {$\Omega$}-tube into the solar
  atmosphere}}.
\bjtitle{Astrophys. J. Lett.}
\bvolume{554},
\bfpage{111}\,--\,\blpage{114}.
doi:\doiurl{10.1086/320935}.
\end{barticle}
\endbibitem

\bibitem[\protect\citeauthoryear{{Fan} and {Gibson}}{2003}]{fan2003}
\begin{barticle}
\bauthor{\bsnm{{Fan}}, \binits{Y.}}, \bauthor{\bsnm{{Gibson}}, \binits{S.E.}}:
\byear{2003},
\batitle{{The emergence of a twisted magnetic flux tube into a preexisting
  coronal arcade}}.
\bjtitle{Astrophys. J. Lett.}
\bvolume{589},
\bfpage{105}\,--\,\blpage{108}.
doi:\doiurl{10.1086/375834}.
\end{barticle}
\endbibitem

\bibitem[\protect\citeauthoryear{{Fan} and {Gibson}}{2004}]{fan2004}
\begin{barticle}
\bauthor{\bsnm{{Fan}}, \binits{Y.}}, \bauthor{\bsnm{{Gibson}}, \binits{S.E.}}:
\byear{2004},
\batitle{{Numerical simulations of three-dimensional coronal magnetic fields
  resulting from the emergence of twisted magnetic flux tubes}}.
\bjtitle{\apj}
\bvolume{609},
\bfpage{1123}\,--\,\blpage{1133}.
doi:\doiurl{10.1086/421238}.
\end{barticle}
\endbibitem

\bibitem[\protect\citeauthoryear{{Fedun} \textit{et~al.}}{2011}]{fedun2011}
\begin{botherref}
\oauthor{\bsnm{{Fedun}}, \binits{V.}}, \oauthor{\bsnm{{Shelyag}}, \binits{S.}},
  \oauthor{\bsnm{{Verth}}, \binits{G.}}, \oauthor{\bsnm{{Mathioudakis}},
  \binits{M.}}, \oauthor{\bsnm{{Erd\'elyi}}, \binits{R.}}:
2011,
{Brief anatomy of MHD waves generated by high-frequency photospheric swirly
  motions}.
\textit{Annales Geophysicae},
submitted.
\end{botherref}
\endbibitem

\bibitem[\protect\citeauthoryear{{Filippov} and {Den}}{2001}]{boris2001}
\begin{barticle}
\bauthor{\bsnm{{Filippov}}, \binits{B.P.}}, \bauthor{\bsnm{{Den}},
  \binits{O.G.}}:
\byear{2001},
\batitle{{A critical height of quiescent prominences before eruption}}.
\bjtitle{\jgr}
\bvolume{106},
\bfpage{25177}\,--\,\blpage{25184}.
doi:\doiurl{10.1029/2000JA004002}.
\end{barticle}
\endbibitem

\bibitem[\protect\citeauthoryear{{Fisher} and {Welsch}}{2008}]{fisher2008}
\begin{botherref}
\oauthor{\bsnm{{Fisher}}, \binits{G.H.}}, \oauthor{\bsnm{{Welsch}},
  \binits{B.T.}}:
2008,
{FLCT: A fast, efficient method for performing local correlation tracking}.
In: {R.~Howe, R.~W.~Komm, K.~S.~Balasubramaniam, \& G.~J.~D.~Petrie } (ed.)
\textit{Subsurface and Atmospheric Influences on Solar Activity},
\textit{Astronomical Society of the Pacific Conference Series}
\textbf{383},
373\,--\,380.
\end{botherref}
\endbibitem

\bibitem[\protect\citeauthoryear{{Golub} \textit{et~al.}}{2007}]{golub2007}
\begin{barticle}
\bauthor{\bsnm{{Golub}}, \binits{L.}}, \bauthor{\bsnm{{Deluca}}, \binits{E.}},
  \bauthor{\bsnm{{Austin}}, \binits{G.}}, \bauthor{\bsnm{{Bookbinder}},
  \binits{J.}}, \bauthor{\bsnm{{Caldwell}}, \binits{D.}},
  \bauthor{\bsnm{{Cheimets}}, \binits{P.}}, \bauthor{\bsnm{{Cirtain}},
  \binits{J.}}, \bauthor{\bsnm{{Cosmo}}, \binits{M.}}, \bauthor{\bsnm{{Reid}},
  \binits{P.}}, \bauthor{\bsnm{{Sette}}, \binits{A.}}, \bauthor{\bsnm{{Weber}},
  \binits{M.}}, \bauthor{\bsnm{{Sakao}}, \binits{T.}}, \bauthor{\bsnm{{Kano}},
  \binits{R.}}, \bauthor{\bsnm{{Shibasaki}}, \binits{K.}},
  \bauthor{\bsnm{{Hara}}, \binits{H.}}, \bauthor{\bsnm{{Tsuneta}},
  \binits{S.}}, \bauthor{\bsnm{{Kumagai}}, \binits{K.}},
  \bauthor{\bsnm{{Tamura}}, \binits{T.}}, \bauthor{\bsnm{{Shimojo}},
  \binits{M.}}, \bauthor{\bsnm{{McCracken}}, \binits{J.}},
  \bauthor{\bsnm{{Carpenter}}, \binits{J.}}, \bauthor{\bsnm{{Haight}},
  \binits{H.}}, \bauthor{\bsnm{{Siler}}, \binits{R.}},
  \bauthor{\bsnm{{Wright}}, \binits{E.}}, \bauthor{\bsnm{{Tucker}},
  \binits{J.}}, \bauthor{\bsnm{{Rutledge}}, \binits{H.}},
  \bauthor{\bsnm{{Barbera}}, \binits{M.}}, \bauthor{\bsnm{{Peres}},
  \binits{G.}}, \bauthor{\bsnm{{Varisco}}, \binits{S.}}:
\byear{2007},
\batitle{{The X-Ray telescope (XRT) for the Hinode mission}}.
\bjtitle{\solphys}
\bvolume{243},
\bfpage{63}\,--\,\blpage{86}.
doi:\doiurl{10.1007/s11207-007-0182-1}.
\end{barticle}
\endbibitem

\bibitem[\protect\citeauthoryear{{Graham} \textit{et~al.}}{2003}]{graham2003}
\begin{botherref}
\oauthor{\bsnm{{Graham}}, \binits{J.D.}}, \oauthor{\bsnm{{Norton}},
  \binits{A.}}, \oauthor{\bsnm{{L{\'o}pez Ariste}}, \binits{A.}},
  \oauthor{\bsnm{{Lites}}, \binits{B.}}, \oauthor{\bsnm{{Socas-Navarro}},
  \binits{H.}}, \oauthor{\bsnm{{Tomczyk}}, \binits{S.}}:
2003,
{The Helioseismic and Magnetic Imager (HMI) on SDO: Full vector magnetography
  with a filtergraph polarimeter}.
In: {J.~Trujillo-Bueno \& J.~Sanchez Almeida} (ed.)
\textit{Astronomical Society of the Pacific Conference Series},
\textit{Astronomical Society of the Pacific Conference Series}
\textbf{307},
131\,--\,136.
\end{botherref}
\endbibitem

\bibitem[\protect\citeauthoryear{{Handy} \textit{et~al.}}{1999}]{handy1999}
\begin{barticle}
\bauthor{\bsnm{{Handy}}, \binits{B.N.}}, \bauthor{\bsnm{{Acton}},
  \binits{L.W.}}, \bauthor{\bsnm{{Kankelborg}}, \binits{C.C.}},
  \bauthor{\bsnm{{Wolfson}}, \binits{C.J.}}, \bauthor{\bsnm{{Akin}},
  \binits{D.J.}}, \bauthor{\bsnm{{Bruner}}, \binits{M.E.}},
  \bauthor{\bsnm{{Caravalho}}, \binits{R.}}, \bauthor{\bsnm{{Catura}},
  \binits{R.C.}}, \bauthor{\bsnm{{Chevalier}}, \binits{R.}},
  \bauthor{\bsnm{{Duncan}}, \binits{D.W.}}, \bauthor{\bsnm{{Edwards}},
  \binits{C.G.}}, \bauthor{\bsnm{{Feinstein}}, \binits{C.N.}},
  \bauthor{\bsnm{{Freeland}}, \binits{S.L.}}, \bauthor{\bsnm{{Friedlaender}},
  \binits{F.M.}}, \bauthor{\bsnm{{Hoffmann}}, \binits{C.H.}},
  \bauthor{\bsnm{{Hurlburt}}, \binits{N.E.}}, \bauthor{\bsnm{{Jurcevich}},
  \binits{B.K.}}, \bauthor{\bsnm{{Katz}}, \binits{N.L.}},
  \bauthor{\bsnm{{Kelly}}, \binits{G.A.}}, \bauthor{\bsnm{{Lemen}},
  \binits{J.R.}}, \bauthor{\bsnm{{Levay}}, \binits{M.}},
  \bauthor{\bsnm{{Lindgren}}, \binits{R.W.}}, \bauthor{\bsnm{{Mathur}},
  \binits{D.P.}}, \bauthor{\bsnm{{Meyer}}, \binits{S.B.}},
  \bauthor{\bsnm{{Morrison}}, \binits{S.J.}}, \bauthor{\bsnm{{Morrison}},
  \binits{M.D.}}, \bauthor{\bsnm{{Nightingale}}, \binits{R.W.}},
  \bauthor{\bsnm{{Pope}}, \binits{T.P.}}, \bauthor{\bsnm{{Rehse}},
  \binits{R.A.}}, \bauthor{\bsnm{{Schrijver}}, \binits{C.J.}},
  \bauthor{\bsnm{{Shine}}, \binits{R.A.}}, \bauthor{\bsnm{{Shing}},
  \binits{L.}}, \bauthor{\bsnm{{Strong}}, \binits{K.T.}},
  \bauthor{\bsnm{{Tarbell}}, \binits{T.D.}}, \bauthor{\bsnm{{Title}},
  \binits{A.M.}}, \bauthor{\bsnm{{Torgerson}}, \binits{D.D.}},
  \bauthor{\bsnm{{Golub}}, \binits{L.}}, \bauthor{\bsnm{{Bookbinder}},
  \binits{J.A.}}, \bauthor{\bsnm{{Caldwell}}, \binits{D.}},
  \bauthor{\bsnm{{Cheimets}}, \binits{P.N.}}, \bauthor{\bsnm{{Davis}},
  \binits{W.N.}}, \bauthor{\bsnm{{Deluca}}, \binits{E.E.}},
  \bauthor{\bsnm{{McMullen}}, \binits{R.A.}}, \bauthor{\bsnm{{Warren}},
  \binits{H.P.}}, \bauthor{\bsnm{{Amato}}, \binits{D.}},
  \bauthor{\bsnm{{Fisher}}, \binits{R.}}, \bauthor{\bsnm{{Maldonado}},
  \binits{H.}}, \bauthor{\bsnm{{Parkinson}}, \binits{C.}}:
\byear{1999},
\batitle{{The transition region and coronal explorer}}.
\bjtitle{\solphys}
\bvolume{187},
\bfpage{229}\,--\,\blpage{260}.
doi:\doiurl{10.1023/A:1005166902804}.
\end{barticle}
\endbibitem

\bibitem[\protect\citeauthoryear{{Heyvaerts}, {Priest}, and
  {Rust}}{1977}]{hey1977}
\begin{barticle}
\bauthor{\bsnm{{Heyvaerts}}, \binits{J.}}, \bauthor{\bsnm{{Priest}},
  \binits{E.R.}}, \bauthor{\bsnm{{Rust}}, \binits{D.M.}}:
\byear{1977},
\batitle{{An emerging flux model for the solar flare phenomenon}}.
\bjtitle{\apj}
\bvolume{216},
\bfpage{123}\,--\,\blpage{137}.
doi:\doiurl{10.1086/155453}.
\end{barticle}
\endbibitem

\bibitem[\protect\citeauthoryear{{Jess} \textit{et~al.}}{2010}]{jess2010}
\begin{barticle}
\bauthor{\bsnm{{Jess}}, \binits{D.B.}}, \bauthor{\bsnm{{Mathioudakis}},
  \binits{M.}}, \bauthor{\bsnm{{Christian}}, \binits{D.J.}},
  \bauthor{\bsnm{{Keenan}}, \binits{F.P.}}, \bauthor{\bsnm{{Ryans}},
  \binits{R.S.I.}}, \bauthor{\bsnm{{Crockett}}, \binits{P.J.}}:
\byear{2010},
\batitle{{ROSA: A high-cadence, synchronized multi-camera solar imaging
  system}}.
\bjtitle{\solphys}
\bvolume{261},
\bfpage{363}\,--\,\blpage{373}.
doi:\doiurl{10.1007/s11207-009-9500-0}.
\end{barticle}
\endbibitem

\bibitem[\protect\citeauthoryear{{Ji} \textit{et~al.}}{2003}]{ji2003}
\begin{barticle}
\bauthor{\bsnm{{Ji}}, \binits{H.}}, \bauthor{\bsnm{{Wang}}, \binits{H.}},
  \bauthor{\bsnm{{Schmahl}}, \binits{E.J.}}, \bauthor{\bsnm{{Moon}},
  \binits{Y.}}, \bauthor{\bsnm{{Jiang}}, \binits{Y.}}:
\byear{2003},
\batitle{{Observations of the failed eruption of a filament}}.
\bjtitle{Astrophys. J. Lett.}
\bvolume{595},
\bfpage{135}\,--\,\blpage{138}.
doi:\doiurl{10.1086/378178}.
\end{barticle}
\endbibitem

\bibitem[\protect\citeauthoryear{{Karpen} and {Boris}}{1986}]{karpen1986}
\begin{barticle}
\bauthor{\bsnm{{Karpen}}, \binits{J.T.}}, \bauthor{\bsnm{{Boris}},
  \binits{J.P.}}:
\byear{1986},
\batitle{{Response of an emerging flux tube to a current-driven instability}}.
\bjtitle{\apj}
\bvolume{307},
\bfpage{826}\,--\,\blpage{837}.
doi:\doiurl{10.1086/164469}.
\end{barticle}
\endbibitem

\bibitem[\protect\citeauthoryear{{Kumar}, {Manoharan}, and
  {Uddin}}{2010a}]{pankaj2010a}
\begin{barticle}
\bauthor{\bsnm{{Kumar}}, \binits{P.}}, \bauthor{\bsnm{{Manoharan}},
  \binits{P.K.}}, \bauthor{\bsnm{{Uddin}}, \binits{W.}}:
\byear{2010}a,
\batitle{{Evolution of solar magnetic field and associated multiwavelength
  phenomena: flare events on 2003 November 20}}.
\bjtitle{\apj}
\bvolume{710},
\bfpage{1195}\,--\,\blpage{1204}.
doi:\doiurl{10.1088/0004-637X/710/2/1195}.
\end{barticle}
\endbibitem

\bibitem[\protect\citeauthoryear{{Kumar} \textit{et~al.}}{2010b}]{pankaj2010c}
\begin{barticle}
\bauthor{\bsnm{{Kumar}}, \binits{P.}}, \bauthor{\bsnm{{Srivastava}},
  \binits{A.K.}}, \bauthor{\bsnm{{Somov}}, \binits{B.V.}},
  \bauthor{\bsnm{{Manoharan}}, \binits{P.K.}}, \bauthor{\bsnm{{Erd{\'e}lyi}},
  \binits{R.}}, \bauthor{\bsnm{{Uddin}}, \binits{W.}}:
\byear{2010}b,
\batitle{{Evidence of solar flare triggering due to loop-loop interaction
  caused by footpoint shear motion}}.
\bjtitle{\apj}
\bvolume{723},
\bfpage{1651}\,--\,\blpage{1664}.
doi:\doiurl{10.1088/0004-637X/723/2/1651}.
\end{barticle}
\endbibitem

\bibitem[\protect\citeauthoryear{{Kumar} \textit{et~al.}}{2010c}]{pankaj2010b}
\begin{barticle}
\bauthor{\bsnm{{Kumar}}, \binits{P.}}, \bauthor{\bsnm{{Srivastava}},
  \binits{A.K.}}, \bauthor{\bsnm{{Filippov}}, \binits{B.}},
  \bauthor{\bsnm{{Uddin}}, \binits{W.}}:
\byear{2010}c,
\batitle{{Multiwavelength Study of the M8.9/3B Solar Flare from AR NOAA
  10960}}.
\bjtitle{\solphys}
\bvolume{266},
\bfpage{39}\,--\,\blpage{58}.
doi:\doiurl{10.1007/s11207-010-9586-4}.
\end{barticle}
\endbibitem

\bibitem[\protect\citeauthoryear{{Lemen} \textit{et~al.}}{2011}]{lemen2011}
\begin{botherref}
\oauthor{\bsnm{{Lemen}}, \binits{J.R.}}, \oauthor{\bsnm{{Title}},
  \binits{A.M.}}, \oauthor{\bsnm{{Akin}}, \binits{D.J.}},
  \oauthor{\bsnm{{Boerner}}, \binits{P.F.}}, \oauthor{\bsnm{{Chou}},
  \binits{C.}}, \oauthor{\bsnm{{Drake}}, \binits{J.F.}},
  \oauthor{\bsnm{{Duncan}}, \binits{D.W.}}, \oauthor{\bsnm{{Edwards}},
  \binits{C.G.}}, \oauthor{\bsnm{{Friedlaender}}, \binits{F.M.}},
  \oauthor{\bsnm{{Heyman}}, \binits{G.F.}}, \oauthor{\bsnm{{Hurlburt}},
  \binits{N.E.}}, \oauthor{\bsnm{{Katz}}, \binits{N.L.}},
  \oauthor{\bsnm{{Kushner}}, \binits{G.D.}}, \oauthor{\bsnm{{Levay}},
  \binits{M.}}, \oauthor{\bsnm{{Lindgren}}, \binits{R.W.}},
  \oauthor{\bsnm{{Mathur}}, \binits{D.P.}}, \oauthor{\bsnm{{McFeaters}},
  \binits{E.L.}}, \oauthor{\bsnm{{Mitchell}}, \binits{S.}},
  \oauthor{\bsnm{{Rehse}}, \binits{R.A.}}, \oauthor{\bsnm{{Schrijver}},
  \binits{C.J.}}, \oauthor{\bsnm{{Springer}}, \binits{L.A.}},
  \oauthor{\bsnm{{Stern}}, \binits{R.A.}}, \oauthor{\bsnm{{Tarbell}},
  \binits{T.D.}}, \oauthor{\bsnm{{Wuelser}}, \binits{J.P.}},
  \oauthor{\bsnm{{Wolfson}}, \binits{C.J.}}, \oauthor{\bsnm{{Yanari}},
  \binits{C.}}, \oauthor{\bsnm{{Bookbinder}}, \binits{J.A.}},
  \oauthor{\bsnm{{Cheimets}}, \binits{P.N.}}, \oauthor{\bsnm{{Caldwell}},
  \binits{D.}}, \oauthor{\bsnm{{Deluca}}, \binits{E.E.}},
  \oauthor{\bsnm{{Gates}}, \binits{R.}}, \oauthor{\bsnm{{Golub}}, \binits{L.}},
  \oauthor{\bsnm{{Park}}, \binits{S.}}, \oauthor{\bsnm{{Podgorski}},
  \binits{W.A.}}, \oauthor{\bsnm{{Bush}}, \binits{R.I.}},
  \oauthor{\bsnm{{Scherrer}}, \binits{P.H.}}, \oauthor{\bsnm{{Gummin}},
  \binits{M.A.}}, \oauthor{\bsnm{{Smith}}, \binits{P.}},
  \oauthor{\bsnm{{Auker}}, \binits{G.}}, \oauthor{\bsnm{{Jerram}},
  \binits{P.}}, \oauthor{\bsnm{{Pool}}, \binits{P.}}, \oauthor{\bsnm{{Soufli}},
  \binits{R.}}, \oauthor{\bsnm{{Windt}}, \binits{D.L.}},
  \oauthor{\bsnm{{Beardsley}}, \binits{S.}}, \oauthor{\bsnm{{Clapp}},
  \binits{M.}}, \oauthor{\bsnm{{Lang}}, \binits{J.}},
  \oauthor{\bsnm{{Waltham}}, \binits{N.}}:
2011,
{The Atmospheric Imaging Assembly (AIA) on the Solar Dynamics Observatory
  (SDO)}.
\textit{\solphys},
in press.
doi:\doiurl{10.1007/s11207-011-9776-8}.
\end{botherref}
\endbibitem

\bibitem[\protect\citeauthoryear{{Liu} \textit{et~al.}}{2009}]{liu2009}
\begin{barticle}
\bauthor{\bsnm{{Liu}}, \binits{Y.}}, \bauthor{\bsnm{{Su}}, \binits{J.}},
  \bauthor{\bsnm{{Xu}}, \binits{Z.}}, \bauthor{\bsnm{{Lin}}, \binits{H.}},
  \bauthor{\bsnm{{Shibata}}, \binits{K.}}, \bauthor{\bsnm{{Kurokawa}},
  \binits{H.}}:
\byear{2009},
\batitle{{New Observation of failed filament eruptions: The influence of
  asymmetric coronal background fields on solar eruptions}}.
\bjtitle{Astrophys. J. Lett.}
\bvolume{696},
\bfpage{70}\,--\,\blpage{73}.
doi:\doiurl{10.1088/0004-637X/696/1/L70}.
\end{barticle}
\endbibitem

\bibitem[\protect\citeauthoryear{{Martin} \textit{et~al.}}{1984}]{martin1984}
\begin{barticle}
\bauthor{\bsnm{{Martin}}, \binits{S.F.}}, \bauthor{\bsnm{{Bentley}},
  \binits{R.D.}}, \bauthor{\bsnm{{Schadee}}, \binits{A.}},
  \bauthor{\bsnm{{Antalova}}, \binits{A.}}, \bauthor{\bsnm{{Kucera}},
  \binits{A.}}, \bauthor{\bsnm{{Dezso}}, \binits{L.}},
  \bauthor{\bsnm{{Gesztelyi}}, \binits{L.}}, \bauthor{\bsnm{{Harvey}},
  \binits{K.L.}}, \bauthor{\bsnm{{Jones}}, \binits{H.}},
  \bauthor{\bsnm{{Livi}}, \binits{S.H.B.}}:
\byear{1984},
\batitle{{Relationships of a growing magnetic flux region to flares}}.
\bjtitle{Adv. Space Res.}
\bvolume{4},
\bfpage{61}\,--\,\blpage{70}.
doi:\doiurl{10.1016/0273-1177(84)90161-3}.
\end{barticle}
\endbibitem

\bibitem[\protect\citeauthoryear{{Okamoto}, {Tsuneta}, and
  {Berger}}{2010}]{okamoto2010}
\begin{barticle}
\bauthor{\bsnm{{Okamoto}}, \binits{T.J.}}, \bauthor{\bsnm{{Tsuneta}},
  \binits{S.}}, \bauthor{\bsnm{{Berger}}, \binits{T.E.}}:
\byear{2010},
\batitle{{A rising cool column as a signature of helical flux emergence and
  formation of prominence and coronal cavity}}.
\bjtitle{\apj}
\bvolume{719},
\bfpage{583}\,--\,\blpage{590}.
doi:\doiurl{10.1088/0004-637X/719/1/583}.
\end{barticle}
\endbibitem

\bibitem[\protect\citeauthoryear{{Okamoto} \textit{et~al.}}{2008}]{okamoto2008}
\begin{barticle}
\bauthor{\bsnm{{Okamoto}}, \binits{T.J.}}, \bauthor{\bsnm{{Tsuneta}},
  \binits{S.}}, \bauthor{\bsnm{{Lites}}, \binits{B.W.}},
  \bauthor{\bsnm{{Kubo}}, \binits{M.}}, \bauthor{\bsnm{{Yokoyama}},
  \binits{T.}}, \bauthor{\bsnm{{Berger}}, \binits{T.E.}},
  \bauthor{\bsnm{{Ichimoto}}, \binits{K.}}, \bauthor{\bsnm{{Katsukawa}},
  \binits{Y.}}, \bauthor{\bsnm{{Nagata}}, \binits{S.}},
  \bauthor{\bsnm{{Shibata}}, \binits{K.}}, \bauthor{\bsnm{{Shimizu}},
  \binits{T.}}, \bauthor{\bsnm{{Shine}}, \binits{R.A.}},
  \bauthor{\bsnm{{Suematsu}}, \binits{Y.}}, \bauthor{\bsnm{{Tarbell}},
  \binits{T.D.}}, \bauthor{\bsnm{{Title}}, \binits{A.M.}}:
\byear{2008},
\batitle{{Emergence of a helical flux rope under an active region prominence}}.
\bjtitle{Astrophys. J. Lett.}
\bvolume{673},
\bfpage{215}\,--\,\blpage{218}.
doi:\doiurl{10.1086/528792}.
\end{barticle}
\endbibitem

\bibitem[\protect\citeauthoryear{{Okamoto} \textit{et~al.}}{2009}]{okamoto2009}
\begin{barticle}
\bauthor{\bsnm{{Okamoto}}, \binits{T.J.}}, \bauthor{\bsnm{{Tsuneta}},
  \binits{S.}}, \bauthor{\bsnm{{Lites}}, \binits{B.W.}},
  \bauthor{\bsnm{{Kubo}}, \binits{M.}}, \bauthor{\bsnm{{Yokoyama}},
  \binits{T.}}, \bauthor{\bsnm{{Berger}}, \binits{T.E.}},
  \bauthor{\bsnm{{Ichimoto}}, \binits{K.}}, \bauthor{\bsnm{{Katsukawa}},
  \binits{Y.}}, \bauthor{\bsnm{{Nagata}}, \binits{S.}},
  \bauthor{\bsnm{{Shibata}}, \binits{K.}}, \bauthor{\bsnm{{Shimizu}},
  \binits{T.}}, \bauthor{\bsnm{{Shine}}, \binits{R.A.}},
  \bauthor{\bsnm{{Suematsu}}, \binits{Y.}}, \bauthor{\bsnm{{Tarbell}},
  \binits{T.D.}}, \bauthor{\bsnm{{Title}}, \binits{A.M.}}:
\byear{2009},
\batitle{{Prominence formation associated with an emerging helical flux rope}}.
\bjtitle{\apj}
\bvolume{697},
\bfpage{913}\,--\,\blpage{922}.
doi:\doiurl{10.1088/0004-637X/697/1/913}.
\end{barticle}
\endbibitem

\bibitem[\protect\citeauthoryear{{Roussev}
  \textit{et~al.}}{2001a}]{roussev2001a}
\begin{barticle}
\bauthor{\bsnm{{Roussev}}, \binits{I.}}, \bauthor{\bsnm{{Galsgaard}},
  \binits{K.}}, \bauthor{\bsnm{{Erd{\'e}lyi}}, \binits{R.}},
  \bauthor{\bsnm{{Doyle}}, \binits{J.G.}}:
\byear{2001}a,
\batitle{{Modelling of explosive events in the solar transition region in a 2D
  environment. I. General reconnection jet dynamics}}.
\bjtitle{\aap}
\bvolume{370},
\bfpage{298}\,--\,\blpage{310}.
doi:\doiurl{10.1051/0004-6361:20010207}.
\end{barticle}
\endbibitem

\bibitem[\protect\citeauthoryear{{Roussev}
  \textit{et~al.}}{2001b}]{roussev2001b}
\begin{barticle}
\bauthor{\bsnm{{Roussev}}, \binits{I.}}, \bauthor{\bsnm{{Galsgaard}},
  \binits{K.}}, \bauthor{\bsnm{{Erd{\'e}lyi}}, \binits{R.}},
  \bauthor{\bsnm{{Doyle}}, \binits{J.G.}}:
\byear{2001}b,
\batitle{{Modelling of explosive events in the solar transition region in a 2D
  environment. II. Various MHD experiments}}.
\bjtitle{\aap}
\bvolume{375},
\bfpage{228}\,--\,\blpage{242}.
doi:\doiurl{10.1051/0004-6361:20010765}.
\end{barticle}
\endbibitem

\bibitem[\protect\citeauthoryear{{Roussev}
  \textit{et~al.}}{2001c}]{roussev2001c}
\begin{barticle}
\bauthor{\bsnm{{Roussev}}, \binits{I.}}, \bauthor{\bsnm{{Doyle}},
  \binits{J.G.}}, \bauthor{\bsnm{{Galsgaard}}, \binits{K.}},
  \bauthor{\bsnm{{Erd{\'e}lyi}}, \binits{R.}}:
\byear{2001}c,
\batitle{{Modelling of solar explosive events in 2D environments. III.
  Observable consequences}}.
\bjtitle{\aap}
\bvolume{380},
\bfpage{719}\,--\,\blpage{726}.
doi:\doiurl{10.1051/0004-6361:20011497}.
\end{barticle}
\endbibitem

\bibitem[\protect\citeauthoryear{{Scharmer}
  \textit{et~al.}}{2003}]{scharmer2003}
\begin{botherref}
\oauthor{\bsnm{{Scharmer}}, \binits{G.B.}}, \oauthor{\bsnm{{Bjelksjo}},
  \binits{K.}}, \oauthor{\bsnm{{Korhonen}}, \binits{T.K.}},
  \oauthor{\bsnm{{Lindberg}}, \binits{B.}}, \oauthor{\bsnm{{Petterson}},
  \binits{B.}}:
2003,
{The 1-meter Swedish solar telescope}.
In: {S.~L.~Keil \& S.~V.~Avakyan} (ed.)
\textit{Society of Photo-Optical Instrumentation Engineers (SPIE) Conference
  Series},
\textit{Society of Photo-Optical Instrumentation Engineers (SPIE) Conference
  Series}
\textbf{4853},
341\,--\,350.
doi:\doiurl{10.1117/12.460377}.
\end{botherref}
\endbibitem

\bibitem[\protect\citeauthoryear{{Scherrer} \textit{et~al.}}{1995}]{sch1995}
\begin{barticle}
\bauthor{\bsnm{{Scherrer}}, \binits{P.H.}}, \bauthor{\bsnm{{Bogart}},
  \binits{R.S.}}, \bauthor{\bsnm{{Bush}}, \binits{R.I.}},
  \bauthor{\bsnm{{Hoeksema}}, \binits{J.T.}}, \bauthor{\bsnm{{Kosovichev}},
  \binits{A.G.}}, \bauthor{\bsnm{{Schou}}, \binits{J.}},
  \bauthor{\bsnm{{Rosenberg}}, \binits{W.}}, \bauthor{\bsnm{{Springer}},
  \binits{L.}}, \bauthor{\bsnm{{Tarbell}}, \binits{T.D.}},
  \bauthor{\bsnm{{Title}}, \binits{A.}}, \bauthor{\bsnm{{Wolfson}},
  \binits{C.J.}}, \bauthor{\bsnm{{Zayer}}, \binits{I.}}, \bauthor{\bsnm{{MDI
  Engineering Team}}}:
\byear{1995},
\batitle{{The Solar oscillations investigation - Michelson doppler imager}}.
\bjtitle{\solphys}
\bvolume{162},
\bfpage{129}\,--\,\blpage{188}.
doi:\doiurl{10.1007/BF00733429}.
\end{barticle}
\endbibitem

\bibitem[\protect\citeauthoryear{{Schrijver} and {Title}}{1999}]{sch1999}
\begin{barticle}
\bauthor{\bsnm{{Schrijver}}, \binits{C.J.}}, \bauthor{\bsnm{{Title}},
  \binits{A.M.}}:
\byear{1999},
\batitle{{Active regions losing their moorings by subsurface reconnection}}.
\bjtitle{\solphys}
\bvolume{188},
\bfpage{331}\,--\,\blpage{344}.
doi:\doiurl{10.1023/A:1005281526160}.
\end{barticle}
\endbibitem

\bibitem[\protect\citeauthoryear{Shelyag \textit{et~al.}}{2011}]{shel2011b}
\begin{barticle}
\bauthor{\bsnm{Shelyag}, \binits{S.}}, \bauthor{\bsnm{Fedun}, \binits{V.}},
  \bauthor{\bsnm{Keenan}, \binits{F.P.}}, \bauthor{\bsnm{Erd\'elyi},
  \binits{R.}}, \bauthor{\bsnm{Mathioudakis}, \binits{M.}}:
\byear{2011},
\batitle{Photospheric magnetic vortex structures}.
\bjtitle{Annales Geophysicae}
\bvolume{29},
\bfpage{883}\,--\,\blpage{887}.
doi:\doiurl{10.5194/angeo-29-883-2011}.
\end{barticle}
\endbibitem

\bibitem[\protect\citeauthoryear{{Shelyag} \textit{et~al.}}{2011}]{shel2011a}
\begin{barticle}
\bauthor{\bsnm{{Shelyag}}, \binits{S.}}, \bauthor{\bsnm{{Keys}}, \binits{P.}},
  \bauthor{\bsnm{{Mathioudakis}}, \binits{M.}}, \bauthor{\bsnm{{Keenan}},
  \binits{F.P.}}:
\byear{2011},
\batitle{{Vorticity in the solar photosphere}}.
\bjtitle{\aap}
\bvolume{526},
\bfpage{A5}.
doi:\doiurl{10.1051/0004-6361/201015645}.
\end{barticle}
\endbibitem

\bibitem[\protect\citeauthoryear{{Shen}, {Liu}, and {Liu}}{2011}]{shen2011}
\begin{barticle}
\bauthor{\bsnm{{Shen}}, \binits{Y.D.}}, \bauthor{\bsnm{{Liu}}, \binits{Y.}},
  \bauthor{\bsnm{{Liu}}, \binits{R.}}:
\byear{2011},
\batitle{{A time series of filament eruptions observed by three eyes from
  space: from failed to successful eruptions}}.
\bjtitle{Res. Astron. Astrophys.}
\bvolume{11},
\bfpage{594}\,--\,\blpage{606}.
doi:\doiurl{10.1088/1674-4527/11/5/009}.
\end{barticle}
\endbibitem

\bibitem[\protect\citeauthoryear{{Shibata} \textit{et~al.}}{1990}]{shibata1990}
\begin{barticle}
\bauthor{\bsnm{{Shibata}}, \binits{K.}}, \bauthor{\bsnm{{Nozawa}},
  \binits{S.}}, \bauthor{\bsnm{{Matsumoto}}, \binits{R.}},
  \bauthor{\bsnm{{Sterling}}, \binits{A.C.}}, \bauthor{\bsnm{{Tajima}},
  \binits{T.}}:
\byear{1990},
\batitle{{Emergence of solar magnetic flux from the convection zone into the
  photosphere and chromosphere}}.
\bjtitle{Astrophys. J. Lett.}
\bvolume{351},
\bfpage{25}\,--\,\blpage{28}.
doi:\doiurl{10.1086/185671}.
\end{barticle}
\endbibitem

\bibitem[\protect\citeauthoryear{{Simon} and {Weiss}}{1997}]{simon1997}
\begin{barticle}
\bauthor{\bsnm{{Simon}}, \binits{G.W.}}, \bauthor{\bsnm{{Weiss}},
  \binits{N.O.}}:
\byear{1997},
\batitle{{Kinematic modeling of vortices in the solar photosphere}}.
\bjtitle{\apj}
\bvolume{489},
\bfpage{960}.
doi:\doiurl{10.1086/304800}.
\end{barticle}
\endbibitem

\bibitem[\protect\citeauthoryear{{Srivastava} \textit{et~al.}}{2010}]{sri2010}
\begin{barticle}
\bauthor{\bsnm{{Srivastava}}, \binits{A.K.}}, \bauthor{\bsnm{{Zaqarashvili}},
  \binits{T.V.}}, \bauthor{\bsnm{{Kumar}}, \binits{P.}},
  \bauthor{\bsnm{{Khodachenko}}, \binits{M.L.}}:
\byear{2010},
\batitle{{Observation of kink instability during small B5.0 solar flare on 2007
  June 4}}.
\bjtitle{\apj}
\bvolume{715},
\bfpage{292}\,--\,\blpage{299}.
doi:\doiurl{10.1088/0004-637X/715/1/292}.
\end{barticle}
\endbibitem

\bibitem[\protect\citeauthoryear{{T{\"o}r{\"o}k} and {Kliem}}{2003}]{torok2003}
\begin{barticle}
\bauthor{\bsnm{{T{\"o}r{\"o}k}}, \binits{T.}}, \bauthor{\bsnm{{Kliem}},
  \binits{B.}}:
\byear{2003},
\batitle{{The evolution of twisting coronal magnetic flux tubes}}.
\bjtitle{\aap}
\bvolume{406},
\bfpage{1043}\,--\,\blpage{1059}.
doi:\doiurl{10.1051/0004-6361:20030692}.
\end{barticle}
\endbibitem

\bibitem[\protect\citeauthoryear{{T{\"o}r{\"o}k} and {Kliem}}{2005}]{torok2005}
\begin{barticle}
\bauthor{\bsnm{{T{\"o}r{\"o}k}}, \binits{T.}}, \bauthor{\bsnm{{Kliem}},
  \binits{B.}}:
\byear{2005},
\batitle{{Confined and ejective eruptions of kink-unstable flux ropes}}.
\bjtitle{Astrophys. J. Lett.}
\bvolume{630},
\bfpage{97}\,--\,\blpage{100}.
doi:\doiurl{10.1086/462412}.
\end{barticle}
\endbibitem

\bibitem[\protect\citeauthoryear{{Wedemeyer-B{\"o}hm} and {Rouppe van der
  Voort}}{2009}]{wede2009}
\begin{barticle}
\bauthor{\bsnm{{Wedemeyer-B{\"o}hm}}, \binits{S.}}, \bauthor{\bsnm{{Rouppe van
  der Voort}}, \binits{L.}}:
\byear{2009},
\batitle{{Small-scale swirl events in the quiet Sun chromosphere}}.
\bjtitle{Astron. Astrophys. Lett.}
\bvolume{507},
\bfpage{9}\,--\,\blpage{12}.
doi:\doiurl{10.1051/0004-6361/200913380}.
\end{barticle}
\endbibitem

\bibitem[\protect\citeauthoryear{{Welsch} \textit{et~al.}}{2007}]{welsch2007}
\begin{barticle}
\bauthor{\bsnm{{Welsch}}, \binits{B.T.}}, \bauthor{\bsnm{{Abbett}},
  \binits{W.P.}}, \bauthor{\bsnm{{De Rosa}}, \binits{M.L.}},
  \bauthor{\bsnm{{Fisher}}, \binits{G.H.}}, \bauthor{\bsnm{{Georgoulis}},
  \binits{M.K.}}, \bauthor{\bsnm{{Kusano}}, \binits{K.}},
  \bauthor{\bsnm{{Longcope}}, \binits{D.W.}}, \bauthor{\bsnm{{Ravindra}},
  \binits{B.}}, \bauthor{\bsnm{{Schuck}}, \binits{P.W.}}:
\byear{2007},
\batitle{{Tests and comparisons of velocity-inversion techniques}}.
\bjtitle{\apj}
\bvolume{670},
\bfpage{1434}\,--\,\blpage{1452}.
doi:\doiurl{10.1086/522422}.
\end{barticle}
\endbibitem

\bibitem[\protect\citeauthoryear{{Xu} \textit{et~al.}}{2010}]{xu2010}
\begin{botherref}
\oauthor{\bsnm{{Xu}}, \binits{Y.}}, \oauthor{\bsnm{{Liu}}, \binits{R.}},
  \oauthor{\bsnm{{Jing}}, \binits{J.}}, \oauthor{\bsnm{{Wang}}, \binits{H.}}:
2010,
{Partial eruption and shrinkage of filaments during solar flares in new solar
  cycle}.
In: \textit{American Astronomical Society Meeting Abstracts \#216},
\textit{Bull. Am. Astron. Soc.}
\textbf{41},
901.
\end{botherref}
\endbibitem

\end{thebibliography}
\end{article} 
\end{document}